\documentclass[12pt]{article} % use larger type; default would be 10pt

\usepackage[utf8]{inputenc} % set input encoding (not needed with XeLaTeX)

\usepackage{geometry} % to change the page dimensions
\usepackage{graphicx} % support the \includegraphics command and options
\graphicspath{ {./Figures/} }

\usepackage{booktabs} % for much better looking tables
\usepackage{array} % for better arrays (eg matrices) in maths
\usepackage{paralist} % very flexible & customisable lists (eg. enumerate/itemize, etc.)
\usepackage{verbatim} % adds environment for commenting out blocks of text & for better verbatim
\usepackage{subfig} % make it possible to include more than one captioned figure/table in a single float
\usepackage{bm}
\usepackage{amsmath}
\usepackage{mathtools}
\usepackage{amsthm}
\usepackage{amssymb}
\usepackage{longtable}
\usepackage{undertilde}
\usepackage{xcolor}
\usepackage{fancyvrb}
\usepackage{multirow}
\usepackage{etoolbox}
\usepackage{float}
\usepackage{natbib}
\usepackage{enumitem}

\let\bbordermatrix\bordermatrix
\patchcmd{\bbordermatrix}{8.75}{4.75}{}{}
\patchcmd{\bbordermatrix}{\left(}{\left[}{}{}
\patchcmd{\bbordermatrix}{\right)}{\right]}{}{}

\newtheorem{theorem}{Theorem}
\newtheorem{lemma}{Lemma}

\theoremstyle{definition}

\theoremstyle{remark}

\theoremstyle{plain}

\DeclareMathAlphabet{\mathpzc}{OT1}{pzc}{m}{it}

\usepackage{fancyhdr} % This should be set AFTER setting up the page geometry
\usepackage{sectsty}
\allsectionsfont{\sffamily\mdseries\upshape} % (See the fntguide.pdf for font help)
\usepackage[nottoc,notlof,notlot]{tocbibind} % Put the bibliography in the ToC
\usepackage[titles,subfigure]{tocloft} % Alter the style of the Table of Contents

\newcommand{\ba}{\boldsymbol{a}}

\newcommand{\bc}{\boldsymbol{c}}

\newcommand{\bvarepsilon}{\boldsymbol{\varepsilon}}

\newcommand{\by}{\boldsymbol{y}}
\newcommand{\bX}{\boldsymbol{X}}

\newcommand{\bzero}{\boldsymbol{0}}

\newcommand{\bbeta}{\boldsymbol{\beta}}

\newcommand{\cov}{\text{cov}}

\newcommand{\var}{\text{var}}
\newcommand{\corr}{\text{corr}}
\newcommand{\RSS}{\text{RSS}}
\newcommand{\AIC}{\text{AIC}}
\newcommand{\BIC}{\text{BIC}}
\newcommand{\GIC}{\text{GIC}}

\makeatletter
\newcommand{\vast}{\bBigg@{4}}
\newcommand{\Vast}{\bBigg@{5}}
\makeatother

\graphicspath{ {Figures/} }

\begin{document}

\begin{center}
\textbf{\Large Finite sample properties of the Buckland-Burnham-Augustin confidence interval centered on a model averaged estimator}
\end{center}

%\smallskip

\begin{center}
	{\large Paul Kabaila$^{\text{a,*}}$, A.H. Welsh$^{\text{b}}$ and Christeen Wijethunga$^{\text{a}}$}
\end{center}

\noindent $^{\text{a}}$ \textsl{\small Department of Mathematics and Statistics, La Trobe University, Victoria 3086, Australia}

\noindent $^{\text{b}}$	\textsl{\small Mathematical Sciences Institute,
	The Australian National University, ACT 2601, Australia}

\bigskip

%\maketitle

\begin{abstract}
\noindent We consider the confidence interval centered on a frequentist model averaged estimator that was proposed by \cite{BucklandEtAl1997}.  In the context of a simple testbed situation involving two linear regression models, we derive exact expressions for the confidence interval and then for the coverage and scaled expected length of the confidence interval.  We use these measures to explore the exact finite sample performance of the Buckland-Burnham-Augustin confidence interval.  We also explore the limiting asymptotic case (as the residual degrees of freedom increases) and compare our results for this case to those obtained for the asymptotic coverage of the confidence interval by \cite{HjortClaeskens2003}.
\end{abstract}

\noindent\textit{Keywords:} coverage; model averaged confidence interval; scaled expected length

\vspace{7cm}

\noindent $^*$ Corresponding author. \textsl{E-mail address:} P.Kabaila@latrobe.edu.au

\newpage

\section{Introduction}

\cite{BucklandEtAl1997} proposed a frequentist model averaged estimator of a general scalar parameter that is a weighted average of estimators obtained under different models.  The model weights were constructed by exponentiating the Akaike Information Criterion (AIC), see \citet[pp.~605--606]{BucklandEtAl1997}.  This kind of model weighting has been adopted in much of the later literature (\cite{FletcherDillingham2011, FletcherTurek2011}).  \cite{BucklandEtAl1997} further proposed a standard error for the model averaged estimator and constructed an approximate (Gaussian) confidence interval centered on the model averaged estimator with width determined by the standard error.   The approach of \cite{BucklandEtAl1997} was enthusiastically adopted by \cite{BurnAnderson2002} and seems to be widely used in practice in the ecological literature. 

\citet[Section~4.3]{HjortClaeskens2003} and \citet[Section~7.5.1]{ClaeskensHjort2008} criticised the confidence interval proposed by \cite{BucklandEtAl1997}. In the context of a general regression model, which includes linear regression and logistic regression as particular cases, they showed that the standard error based on formula (9) of \cite{BucklandEtAl1997} is asymptotically incorrect, so that the nominal coverage of the confidence interval is not the actual coverage.  The analyses of \cite{HjortClaeskens2003} and \cite{ClaeskensHjort2008} do not seem to have had much impact in applied fields.  Although the conclusions are clear and hold for very general regression models, the results themselves are complicated, difficult to follow and, as large sample results, deemed not very relevant to practice. 

\cite{KabailaWelshAbeysekera2016} set up a very simple testbed situation for evaluating the exact finite sample frequentist properties of model averaged confidence intervals.  This testbed involves computing a confidence interval by model averaging over two nested linear regression models with unknown error variance, and then computing the coverage probability and scaled expected length properties of this confidence interval.  The scaled expected length is the expected length of the model averaged confidence interval divided by the expected length of the standard confidence interval (with the same minimum coverage probability and for the same parameter) computed under the full model without any model selection.  Its computation gives far more insight than the coverage alone, allowing us for example to see when good coverage is obtained at the expense of  excessive length.  The testbed was used by \cite{KabailaWelshAbeysekera2016}  to evaluate both the model averaged profile likelihood confidence interval of \cite{FletcherTurek2011} and the model averaged tail area confidence interval of \cite{TurekFletcher2012}. This testbed was also used
by \cite{ KabailaWelshMainzer2017} to further evaluate the tail area confidence interval of \cite{TurekFletcher2012}.
 These papers showed that the tail area interval performs quite well provided that we do not put too much weight on the simpler of the two models.  On the other hand, there are situations in which the model averaged profile likelihood intervals are worse than the standard confidence interval used after model selection but ignoring the model selection process.

Our aim is to analyse the exact finite sample properties of the \cite{BucklandEtAl1997} confidence interval using the standard error based on the formula (9) of \cite{BucklandEtAl1997} in the testbed situation of two nested linear regression models with unknown error variance. Specifically, we want to find the exact finite sample coverage probability and scaled expected length properties of this confidence interval.  We also allow the use of different model selection criteria, such as the Bayesian Information Criterion (BIC) instead of AIC, to determine the model averaging weights so that one could explore the effect of changing the weights.  However, the results of \cite{KabailaWelshAbeysekera2016}, \cite{KabailaWelshMainzer2017} and \cite{Kabaila2018}
suggest that BIC weights put too much weight on the simpler model, producing confidence intervals with poorer performance than the AIC weights.

We define the testbed situation and the parametrisation we use in Section \ref{testbed}.  In Section \ref{mavest}, we obtain explicit expressions for the model averaged estimator and the standard error proposed for it by  \cite{BucklandEtAl1997}  in the testbed situation.  These expressions together enable us to obtain an explicit expression for the confidence interval centered on the model averaged  estimator proposed by  \cite{BucklandEtAl1997}   (which we denote $J$) in the testbed situation. In Section \ref{cover}, we then derive expressions for the coverage probability and scaled expected length of the confidence interval $J$ in the testbed situation. We present numerical results for small residual degrees of freedom $m$ under various parameter settings in Section \ref{results}.  We then consider the limiting case as $m \rightarrow \infty$ with the dimension of the regression parameter fixed in Section \ref{largem} and compare these with the large sample results of \cite{HjortClaeskens2003} in Subsection \ref{compare}.  We conclude with a brief discussion in Section \ref{discuss}.

%------------------------------------------------------------------------------------------

\section{Testbed model and parametrisation} \label{testbed}

Our testbed situation involves two nested linear regression models with unknown error variance which we call the full model ${\cal M}_2$, and a simpler model ${\cal M}_1$.  The full model ${\cal M}_2$ is given by
\begin{equation*}
\by = \bX \bbeta + \bvarepsilon,
\end{equation*}
where $\by$ is an $n$-vector of random responses, $\bX$ is a $n \times p$ matrix with known, linearly independent columns, $\bbeta$ is an $p$-vector of unknown  parameters and $\bvarepsilon$ is an $n$-vector of random errors with a $N(\bzero, \sigma^2 \boldsymbol{I})$ distribution in which $\sigma^2$ is an unknown, positive parameter.  We assume throughout that that $n$ and $p$ are given.  Suppose that the parameter of interest is $\theta = \ba^\top \bbeta$, where $\ba$ is a specified $p$-vector ($\ba \neq 0$).  To define the simpler model, we define another parameter $\tau = \bc^\top \bbeta - t$, where the vector $\bc$ and the number $t$ are specified and $\ba$ and $\bc$ are linearly independent. The model ${\cal M}_1$ is ${\cal M}_2$ with $\tau=0$.  

Let $\widehat{\bbeta}=(\bX^\top\bX)^{-1}\bX^\top\by$ denote the least squares estimator of $\bbeta$ and $\widehat{\sigma}^2 = \big(\by - \bX \widehat\bbeta\big)^\top \big(\by - \bX \widehat\bbeta\big) / m$, where $m = n - p$, the usual unbiased estimator of $\sigma^2$.  We set $\widehat{\theta} = \ba^\top \widehat{\bbeta}$ and $\widehat{\tau} = \bc^\top \widehat{\bbeta} - t$. Define the known quantities $v_\theta = \var\big(\widehat{\theta}\big)/\sigma^2 = \ba^\top(\bX^\top\bX)^{-1}\ba$, $v_\tau = \var\big(\widehat{\tau}\big)/\sigma^2= \bc^\top(\bX^\top\bX)^{-1}\bc$ and $\rho = \corr\big(\widehat{\theta}, \widehat{\tau}\big) =  \ba^\top(\bX^\top\bX)^{-1}\bc/\{\ba^\top(\bX^\top\bX)^{-1}\ba \; \bc^\top(\bX^\top\bX)^{-1}\bc\}^{1/2}$. Finally, let 
$\gamma = \tau/ \big( \sigma \, v_\tau^{1/2} \big)$
and 
$\widehat{\gamma} = \widehat{\tau}/ \big( \widehat{\sigma} \, v_\tau^{1/2} \big)$.  Note that $\gamma$ is a measure of the closeness of the models ${\cal M}_1$ and ${\cal M}_2$ and $\widehat{\gamma} $ is an estimator of that measure.

%------------------------------------------------------------------------------------------

\section{The model averaged estimator and its standard deviation} \label{mavest}

\subsection{The model averaged estimator $\widetilde{\theta}$}

Following \citet[p.604]{BucklandEtAl1997}, the model averaged estimator over the class of models  $\{{\cal M}_1, {\cal M}_2\}$ is $\widetilde{\theta} = w_1 \widehat{\theta}_1 + w_2 \widehat{\theta}_2$, where $\widehat{\theta}_1$ and $\widehat{\theta}_2$ are estimators of $\theta$ under the models ${\cal M}_1$ and  ${\cal M}_2$ respectively and $w_1$ and $w_2$, satisfying $w_1 + w_2 = 1$, are the data-based weights for  the models ${\cal M}_1$ and  ${\cal M}_2$ respectively.

We can take $\widehat{\theta}_2 = \widehat{\theta} = \ba^\top \widehat{\bbeta}$. From \citet[p 3421]{KaGi2009a}, 
\begin{align}
\widehat{\theta}_1 &= \widehat{\theta} - \frac{\cov(\widehat{\theta}, \widehat{\tau})}{\var(\widehat{\tau})} \, \widehat{\tau} = \widehat{\theta} - \frac{\cov(\widehat{\theta}, \widehat{\tau})}{\sigma^2 v_\tau} \, \widehat{\tau}= \widehat{\theta} - \frac{\cov(\widehat{\theta}, \widehat{\tau})}{\sigma v_\theta^{1/2} \, \sigma v_\tau^{1/2}} \, v_\theta^{1/2} \frac{\widehat{\tau}}{v_\tau^{1/2}}= \widehat{\theta} - \rho\, \frac{v_\theta^{1/2}\, \widehat{\tau} }{v_\tau^{1/2}}, \label{estimate_theta1}
\end{align}
so we can write
\begin{align}
%\notag
\widetilde{\theta} 
&= (1 - w_1) \widehat{\theta} + w_1 \Big( \widehat{\theta} - \rho\, \frac{v_\theta^{1/2}\, \widehat{\tau} }{ v_\tau^{1/2}} \Big)= \widehat\theta - \rho\, \frac{v_\theta^{1/2}\,\widehat{\tau}}{v_\tau^{1/2}} \, w_1. \label{theta_tilde}
\end{align}

\citet[p.606]{BucklandEtAl1997} defined the model weights to be
\begin{align*}
w_1 &= \frac{\exp \big( -\AIC(1)/2 \big) }{\exp \big( -\AIC(1)/2 \big) + \exp \big( -\AIC(2)/2 \big)} = \frac{1}{1 + \exp \left(\displaystyle \frac{1}{2} \Big( \AIC(1) - \AIC(2) \Big) \right)},
\end{align*}
with $w_2=1-w_1$, where $\AIC(k)$ is  Akaike Information Criterion for model 
${\cal M}_k$.  In a slight generalisation, we replace the Akaike Information Criterion by the Generalized Information Criterion
\begin{equation*}
\GIC(k) = -2 L_k + d\, (p + k - 1),
\end{equation*}
where $L_k$ is the likelihood for model ${\cal M}_k$ ($k = 1,2$), $d = 2$ for $\AIC$ and $d = \ln(n)$ for $\BIC$.  That is, our weights contain the Akaike Information Criterion weights as a special case.  The maximum log-likelihood for model ${\cal M}_k$ is 
\begin{equation*}
L_k = -\frac{n}{2} \ln \left( 2 \pi \, \frac{\RSS_k}{n} \right) - \frac{n}{2},
\end{equation*} 
where $\RSS_k$ denotes the residual sum of squares for model ${\cal M}_k\, (k=1,2)$.  Thus
\begin{align*}
\GIC(k) 
%&= n \ln \left( 2 \pi \, \frac{\RSS_k}{n} \right) + n + d\, (p + k - 1)\\
%&
= \text{ constant } + n\, \ln \left( \frac{\RSS_k}{n} \right) + d\, (p + k -1).
\end{align*}
Now $\text{RSS}_2 = m \, \hat{\sigma}^2$ and, using the results stated by \citeauthor{Graybill1976} (\citeyear{Graybill1976}, p.222), it may be shown that
\begin{equation*}
\RSS_1 
%=  \frac{\widehat{\tau}^2}{v_\tau}+\RSS 
= \frac{\widehat{\tau}^2}{v_\tau} + m \, \hat{\sigma}^2. 
\end{equation*} 
Therefore
\begin{equation*}
\GIC(1) = \text{ constant } + n\, \ln \left( \frac{m\hat{\sigma}^2 + \widehat\tau^2/v_\tau}{n} \right) + d\, p
\end{equation*}
and
\begin{equation*}
\GIC(2) =\text{ constant } +  n\, \ln \left( \frac{m\hat{\sigma}^2}{n} \right) + d\, (p+1).
\end{equation*}
%
%Consequently,
%%
%\begin{align*}
%\GIC(1) - \GIC(2) &= n\, \ln \left( \left(\frac{\widehat{\tau}^2}{v_\tau} + m \widehat\sigma^2 \right) \Big/ n \right) + d\, p -  n\, \ln \left( m \frac{\widehat{\sigma}^2}{n} \right) - d\, (p+1) 
%\\
%&= n\, \ln \left( 1 + \frac{\widehat{\gamma}^2}{m} \right) - d,
%\end{align*}
%%
%and hence
%%
%\begin{align*}
%\exp \left(\frac{1}{2} \Big( \GIC(1) - \GIC(2) \Big) \right) 
%&= \left( 1 + \frac{\widehat{\gamma}^2}{m} \right)^{n/2} \, \exp\big(-d/2 \big)
%\end{align*}
%%
%so
Hence
\begin{equation*}
w_1 =\frac{1}{1 + \exp \left(\displaystyle \frac{1}{2} \Big( \GIC(1) - \GIC(2) \Big) \right)} = \frac{1}{1 + \left( 1 + \displaystyle \frac{\widehat{\gamma}^2}{m} \right)^{n/2} \, \exp\big(-d/2 \big)}.
\end{equation*}
It is convenient to define
\begin{equation}
	\label{w1}
w_1(x) = \frac{1}{1 + \left( 1 + \displaystyle 
	\frac{ x^2}{m} \right)^{n/2} \, \exp\big(-d/2 \big)} = \frac{1}{1 + \left( 1 + \displaystyle 
	\frac{ x^2}{m} \right)^{(m+p)/2} \, \exp\big(-d/2 \big)},
\end{equation}
%, 
where we have written $w_1(x)$ as a function of $x$, $m$, $p$ and $d$. Let $k(x)=x \, w_1(x)$ so we can write (\ref{theta_tilde}) as
\begin{align*}
\widetilde{\theta} =  \widehat\theta - \rho\, v_\theta^{1/2}\, \big(\widehat{\tau} / v_\tau^{1/2} \big) \, 
w_1 \left( \widehat{\gamma} \right) = \widehat\theta - \rho\, v_\theta^{1/2}\, \widehat{\sigma}\widehat{\gamma}\, 
w_1 \left( \widehat{\gamma} \right) = \widehat\theta - \rho\, v_\theta^{1/2}\, \widehat{\sigma} \, k(\widehat{\gamma}).
\end{align*}
%
%Note that
%$k(x)$ is an odd function of $x$.
%

\subsection{The standard error of the model averaged estimator $\widetilde{\theta}$}

We use formula (9) of \cite{BucklandEtAl1997} as the standard deviation of the model averaged estimator $\widetilde{\theta}$:
\begin{equation*}
 \sum_{k=1}^{2} w_k \, \sqrt{\big( \text{variance of $\widehat{\theta}_k$ assuming ${\cal M}_k$ is true} \big) + \Big( E(\widehat{\theta}_k) - \theta \Big)^2 }.
\end{equation*} 
While the derivation of this formula is not very clear, the terms in it are explicit and
\cite{BucklandEtAl1997} also specify an estimator of (9) which we call the standard error of $\widetilde{\theta}$; the estimator of $\big( \text{variance of $\widehat{\theta}_k$ assuming ${\cal M}_k$ is true} \big)$ is obtained in the obvious way, assuming that the model ${\cal M}_k$ is the true model and $\Big( E(\widehat{\theta}_k) - \theta \Big)^2$ is estimated by $\Big( \widehat{\theta}_k - \widetilde{\theta} \Big)^2$.

We have
\begin{align}
\label{var_theta2}
\big( \text{variance of $\widehat{\theta}_2$ assuming ${\cal M}_2$ is true} \big) = \sigma^2 v_\theta.
\end{align}
%
%Assuming that ${\cal M}_1$ is true is equivalent to assuming that $\tau=0$.  Under this assumption, $E\left(\displaystyle \widehat{\theta} - \rho\, v_\theta^{1/2}\, \frac{\widehat{\tau}}{v_\tau^{1/2}} \right) = \theta$, so
%%
%\begin{align}
%\notag
%\var \left( \widehat{\theta} - \rho\, v_\theta^{1/2}\, \frac{\widehat{\tau}}{v_\tau^{1/2}} \right) 
%&= \var(\widehat{\theta}) -  2\rho\, \frac{v_\theta^{1/2}}{v_\tau^{1/2}}\, \cov(\widehat{\theta}, \widehat{\tau}) + \rho^2\, \frac{v_\theta}{v_\tau}\, \var(\widehat{\tau})\\
%\label{var_theta1}
%&= \sigma^2\, v_\theta (1-\rho^2).
%\end{align}  
%%
It may be shown that 
\begin{align}
\big( \text{variance of $\widehat{\theta}_1$ assuming ${\cal M}_1$ is true} \big) 
%&= \var \left( \widehat{\theta} - \rho\, v_\theta^{1/2}\, \frac{\widehat{\tau}}{v_\tau^{1/2}} \right)
%\notag
%\var \left( \widehat{\theta} - \rho\, v_\theta^{1/2}\, \frac{\widehat{\tau}}{v_\tau^{1/2}} \right) 
%&= \var(\widehat{\theta}) -  2\rho\, \frac{v_\theta^{1/2}}{v_\tau^{1/2}}\, \cov(\widehat{\theta}, \widehat{\tau}) + \rho^2\, \frac{v_\theta}{v_\tau}\, \var(\widehat{\tau})\\
\label{var_theta1}
%&
= \sigma^2\, v_\theta \, (1-\rho^2).
\end{align}

\medskip

\noindent The usual estimator of \eqref{var_theta2}, assuming that ${\cal M}_2$ is the true model, is $\widehat{\sigma}^2\, v_\theta$. The usual estimator of \eqref{var_theta1}, assuming that ${\cal M}_1$ is the true model, is
\begin{equation*}
\frac{m \widehat{\sigma}^2 + \big( \widehat{\tau}^2 / v_\tau\big) }{m+1} \, v_\theta\, (1-\rho^2).
\end{equation*} 

Now for the remaining terms,
\begin{align*}
\widehat{\theta}_1 - \widetilde{\theta} &= \widehat{\theta}_1 - \Big( w_1 \widehat\theta_1 + w_2 \widehat\theta_2\Big)= (1-w_1) \widehat\theta_1 - w_2 \widehat{\theta}_2 = w_2 \big( \widehat{\theta}_1 - \widehat{\theta}_2 \big)
\end{align*}
and, similarly,
$\widehat{\theta}_2 - \widetilde{\theta} 
 = w_1 \big( \widehat{\theta}_2 - \widehat{\theta}_1 \big)$.
Hence
\begin{align*}
\Big(\widehat{\theta}_1 - \widetilde{\theta}\Big)^2 &= w_2^2 \big( \widehat{\theta}_1 - \widehat{\theta}_2 \big)^2 \quad \mbox{ and }\quad
\Big(\widehat{\theta}_2 - \widetilde{\theta}\Big)^2 = w_1^2 \big( \widehat{\theta}_1 - \widehat{\theta}_2 \big)^2.
\end{align*}

\medskip

\noindent By \eqref{estimate_theta1},
$\widehat{\theta}_1 - \widehat{\theta}_2 = \widehat{\theta}_1 - \widehat{\theta} 
= - \rho\, v_\theta^{1/2}\, \widehat{\tau} / v_\tau^{1/2}
= - \rho\, \widehat{\sigma} \, v_\theta^{1/2}\, \widehat{\gamma}$. 
Thus
\begin{align*}
\Big(\widehat{\theta}_1 - \widetilde{\theta}\Big)^2 &= w_2^2 \, \rho^2 \, \widehat{\sigma}^2\, v_\theta\, \widehat{\gamma}^2 = (1-w_1)^2 \, \rho^2 \, \widehat{\sigma}^2\, v_\theta\, \widehat{\gamma}^2\\
\Big(\widehat{\theta}_2 - \widetilde{\theta}\Big)^2 &= w_1^2 \, \rho^2 \, \widehat{\sigma}^2\, v_\theta\, \widehat{\gamma}^2.
\end{align*}
Hence, in the testbed situation, the estimator of (9) proposed by \cite{BucklandEtAl1997} is 
\begin{align*}
&w_1\big(\widehat{\gamma} \big) \left( \frac{m \widehat{\sigma}^2 + \big( \widehat{\tau}^2 / v_\tau\big) }{m+1} \, v_\theta\, (1-\rho^2) + \big(1-w_1\big(\widehat{\gamma} \big)\big)^2 \, \rho^2 \, \widehat{\sigma}^2\, v_\theta\, \widehat{\gamma}^2 \right)^{1/2}  
\\
&\qquad \qquad \qquad \qquad \qquad \qquad \qquad \qquad+  \left(1-w_1\big(\widehat{\gamma} \big) \right) \Big( \widehat{\sigma}^2 \, v_\theta + w_1^2\big(\widehat{\gamma} \big) \, \rho^2 \, \widehat{\sigma}^2\, v_\theta\, \widehat{\gamma}^2 \Big)^{1/2}.
\end{align*}
We write this as
$\widehat{\sigma} \, v_\theta^{1/2} \, r \big(\widehat{\gamma} , \rho\big)$, where
\begin{align*}
r(x, \rho) =   w_1(x) \left( \frac{m + x^2}{m+1} (1-\rho^2) + (1-w_1(x))^2 \rho^2\, x^2 \right)^{1/2} + (1-w_1(x)) \Big( 1  + w_1^2(x) \, \rho^2\,  x^2 \Big)^{1/2}.
\end{align*}

\subsection{The confidence interval for $\theta$} 

The confidence interval for $\theta$, centered on  $\widetilde{\theta}$, proposed by \cite{BucklandEtAl1997} is 
\begin{align}
\label{eq:BBA_CI}
\left[\widetilde{\theta} \pm t_{m, 1-\alpha/2} \, \widehat{\sigma} \, v_\theta^{1/2} \, r \big(\widehat{\gamma}, \rho \big)\right]
&= \left[\widehat\theta -  \rho\, \widehat{\sigma} \,v_\theta^{1/2} \, 
k\big(\widehat{\gamma} \big) \,
\pm t_{m, 1-\alpha/2} \, \widehat{\sigma} \, v_\theta^{1/2} \, r \big(\widehat{\gamma}, \rho \big)\right].
\end{align}
We can write
\begin{align*}
\widehat{\gamma} = \widehat{\tau}/\big( \widehat{\sigma} \, v_\tau^{1/2} \big) = \{\widehat{\tau}/\big( \sigma \, v_\tau^{1/2} \big)\}\big(\sigma/ \widehat{\sigma}\big) = \widetilde{\gamma} \big/ W,
\end{align*}
where $\widetilde{\gamma} = \widehat{\tau}/\big( \sigma \, v_\tau^{1/2} \big)$ and
$W = \widehat{\sigma}/\sigma$.   To find convenient formulas for the coverage probability and the expected length of this confidence interval, we express all quantities of interest in terms of  $\big(\widehat{\theta}, \widetilde{\gamma} \big)$ and $W$. 
We denote the pdf of $W$ by $f_W$. 
Note that $\big(\widehat{\theta}, \widetilde{\gamma} \big)$ and $W$ are independent and 
\begin{align}
\label{bivariate_theta_gamma}
\left[ {\begin{array}{c}
	\widehat{\theta} \\
	\widetilde\gamma
	\end{array} } \right] \sim N \left(
\left[ {\begin{array}{c}
	\theta \\
	\gamma
	\end{array} } \right],
\left[ {\begin{array}{cc}
	\sigma^2 \, v_\theta & \rho \, \sigma \, {v_\theta}^{1/2} \\
	\rho \, \sigma \, {v_\theta}^{1/2} & 1
	\end{array} } \right]
\right).
\end{align}
%
%Expressed in terms of $\widehat{\theta}$, $\widetilde{\gamma}$ and $W$,
%%
%\begin{align*}
%\widetilde{\theta} = \widehat{\theta} -\rho \, v_\theta^{1/2} \, \sigma\,  W \,  k\big(\widetilde{\gamma} / W \big).
%\end{align*}
%%
The confidence interval \eqref{eq:BBA_CI}, with nominal coverage $1-\alpha$, is
\begin{equation*}
J = \left[\widehat{\theta} -\rho \, \sigma\,  W \,  v_\theta^{1/2} \, k\big(\widetilde{\gamma} / W \big) 
\pm t_{m, 1-\alpha/2} \, \sigma \, W \,  v_\theta^{1/2} \, r \big(\widetilde{\gamma} / W, \rho \big)\right].
\end{equation*}
%

%------------------------------------------------------------------------------

\section{The coverage probability and scaled expected length of $J$} \label{cover}

\subsection{Coverage probability}

The coverage probability of the confidence interval $J$ for $\theta$ is established in the following theorem.

\begin{theorem}
	\label{Theorem_CP}
	The coverage probability of the confidence interval $J$, with nominal coverage $1-\alpha$,  is a function of $(\gamma, \rho)$ so we denote this coverage probability by  $CP(\gamma, \rho)$. Let 
	%
%	\begin{align*}
%	\ell(\widetilde{\gamma}, W, \rho) 
%	&= \rho \,  W \,  k\big(\widetilde{\gamma} / W \big) - t_{m, 1-\alpha/2} \,   W \, r \big(\widetilde{\gamma} / W, \rho \big)  
%	\\
%	u(\widetilde{\gamma}, W, \rho) 
%	&= \rho \,  W \,  k\big(\widetilde{\gamma} / W \big) + t_{m, 1-\alpha/2} \,   W \, r \big(\widetilde{\gamma} / W, \rho \big).
%	\end{align*}
	%
\begin{align*}
	\ell(\gamma, w, \rho) 
	&= \rho \,  w \,  k\big(\gamma / w \big) - t_{m, 1-\alpha/2} \,   w \, r \big(\gamma / w, \rho \big)  
	\\
	u(\gamma, w, \rho) 
	&= \rho \, w \,  k\big(\gamma / w \big) + t_{m, 1-\alpha/2} \,   w \, r \big(\gamma / w, \rho \big).
\end{align*}
	Then
	%
%	\begin{enumerate}[label=(\alph*)]
%		\item 
%		\begin{equation}
%		\label{CP_final}
%		CP(\gamma, \rho) = \int_{0}^{\infty} \int_{-\infty}^{\infty} \Psi \Big( \ell(y+\gamma, w, \rho), u(y+\gamma, w ,\rho); \, \rho \, y, 1-\rho^2 \Big) \, \phi(y)\, dy\, f_W(w)\, dw,
%		\end{equation}
%	%
%	where $\Psi(a, b; \mu, v) = P(a \leq Z \leq b)$ for $Z \sim N(\mu, v)$. 
%		
%		\item For every given $\rho$, $CP(\gamma, \rho)$ is an even function of $\gamma$ and, for every given $\gamma$, $CP(\gamma, \rho)$ is an even function of $\rho$.
%	\end{enumerate}
\begin{equation}
		\label{CP_final}
CP(\gamma, \rho) = \int_{0}^{\infty} \int_{-\infty}^{\infty} \Psi \Big( \ell(y+\gamma, w, \rho), u(y+\gamma, w ,\rho); \, \rho \, y, 1-\rho^2 \Big) \, \phi(y)\, dy\, f_W(w)\, dw,
\end{equation}
where $\Psi(a, b; \mu, v) = P(a \leq Z \leq b)$ for $Z \sim N(\mu, v)$.  For every given $\rho$, $CP(\gamma, \rho)$ is an even function of $\gamma$ and, for every given $\gamma$, $CP(\gamma, \rho)$ is an even function of $\rho$.

\end{theorem}

\noindent The proof of this result is given in  Appendix \ref{Proof_CP}.  It follows that, for given $m$ and $p$, we are able to describe the coverage probability of $J$ using only the parameters $|\rho|$ and $|\gamma|$.

\subsection{Scaled expected length}

The scaled expected length of the confidence interval $J$ for $\theta$ is established in the following theorem.

\begin{theorem}
	\label{Theorem_SEL}
	The scaled expected length of the confidence interval $J$, with nominal coverage $1-\alpha$,  is a function of $(\gamma, \rho)$ so we denote this scaled expected length by $SEL(\gamma, \rho)$. Then
%	
%	\begin{enumerate}[label=(\alph*)]
%		\item 
%		\begin{align}
%		\notag
%		&SEL(\gamma, \rho) \\
%		\label{SEL_final}
%		&= \frac{t_{m, 1-\alpha/2}}{t_{m, (1+c_{\rm min})/2}}\, \left(\frac{m}{2}\right)^{1/2} \frac{\Gamma(m/2)}{\Gamma\big((m+1)/2\big)}\, \int_{0}^{\infty} \int_{-\infty}^{\infty} w\, r\left(\frac{y+\gamma}{w}, \rho \right)\, \phi(y)\, dy \, f_W(w)\, dw.
%		\end{align}
%		
%		\item For every given $\rho$, $SEL(\gamma, \rho)$ is an even function of $\gamma$ and, for every given $\gamma$, $SEL(\gamma, \rho)$ is an even function of $\rho$.
%	\end{enumerate}
\begin{align}
		\notag
		&SEL(\gamma, \rho) \\
		\label{SEL_final}
		&= \frac{t_{m, 1-\alpha/2}}{t_{m, (1+c_{\rm min})/2}}\, \left(\frac{m}{2}\right)^{1/2} \frac{\Gamma(m/2)}{\Gamma\big((m+1)/2\big)}\, \int_{0}^{\infty} \int_{-\infty}^{\infty} w\, r\left(\frac{y+\gamma}{w}, \rho \right)\, \phi(y)\, dy \, f_W(w)\, dw.
\end{align}
For every given $\rho$, $SEL(\gamma, \rho)$ is an even function of $\gamma$ and, for every given $\gamma$, $SEL(\gamma, \rho)$ is an even function of $\rho$.
	
\end{theorem}

\noindent The proof  of this result is given in Appendix \ref{Proof_SEL}.  It follows that, for given $m$ and $p$, we are able to describe the scaled expected length using only the parameters $|\rho|$ and $|\gamma|$.

%-----------------------------------------------------------------------

\section{Numerical results for small $\boldsymbol{m}$}
\label{results}

We focus on the properties of the confidence interval $J$, with nominal coverage 0.95, computed using AIC weights ($d=2)$.  We constructed a number of plots of the coverage probability and scaled expected length of $J$ against $|\gamma|$ for different values of $m \in \{ 1,2,3,10 \}$, $p \in \{ 3,6,12,124 \}$ and $|\rho| \in \{ 0.2,0.5,0.7,0.9 \}$.  Some explanation of how we carried out the calculations for these plots is included in the Supplementary Material.  We present here a selection of these plots; additional plots are included in the Supplementary Material.  Recall that in (\ref{w1}), we expressed the weights used in the model averaged estimator in terms of $m$ and $p$; since $m=n-p$, we can use any pair of $m$, $p$ and $n=m+p$ and we choose to use $m$ and $p$. 

Generally, the plots show that coverage of $J$ approaches to the nominal level as $|\gamma|$ increases whereas the scaled expected length of $J$ does not necessarily approach $1$ (as we might hope) as $|\gamma|$ increases.  The minimum coverage probability of $J$ is a decreasing continuous function of $|\rho|$. Also, the minimum coverage probability of $J$ is a decreasing continuous function of $m$. When $m=1$, the coverage probability is extremely close to the nominal coverage for any given $|\rho|$ %(see for example Figures \ref{CP_AIC_m_1_rho_2_5_p3} and \ref{CP_AIC_m_1_rho_7_9_p3})
and decreases as $m$ increases.

\begin{figure}[H]
	\centering
	\includegraphics[width=0.9\textwidth]{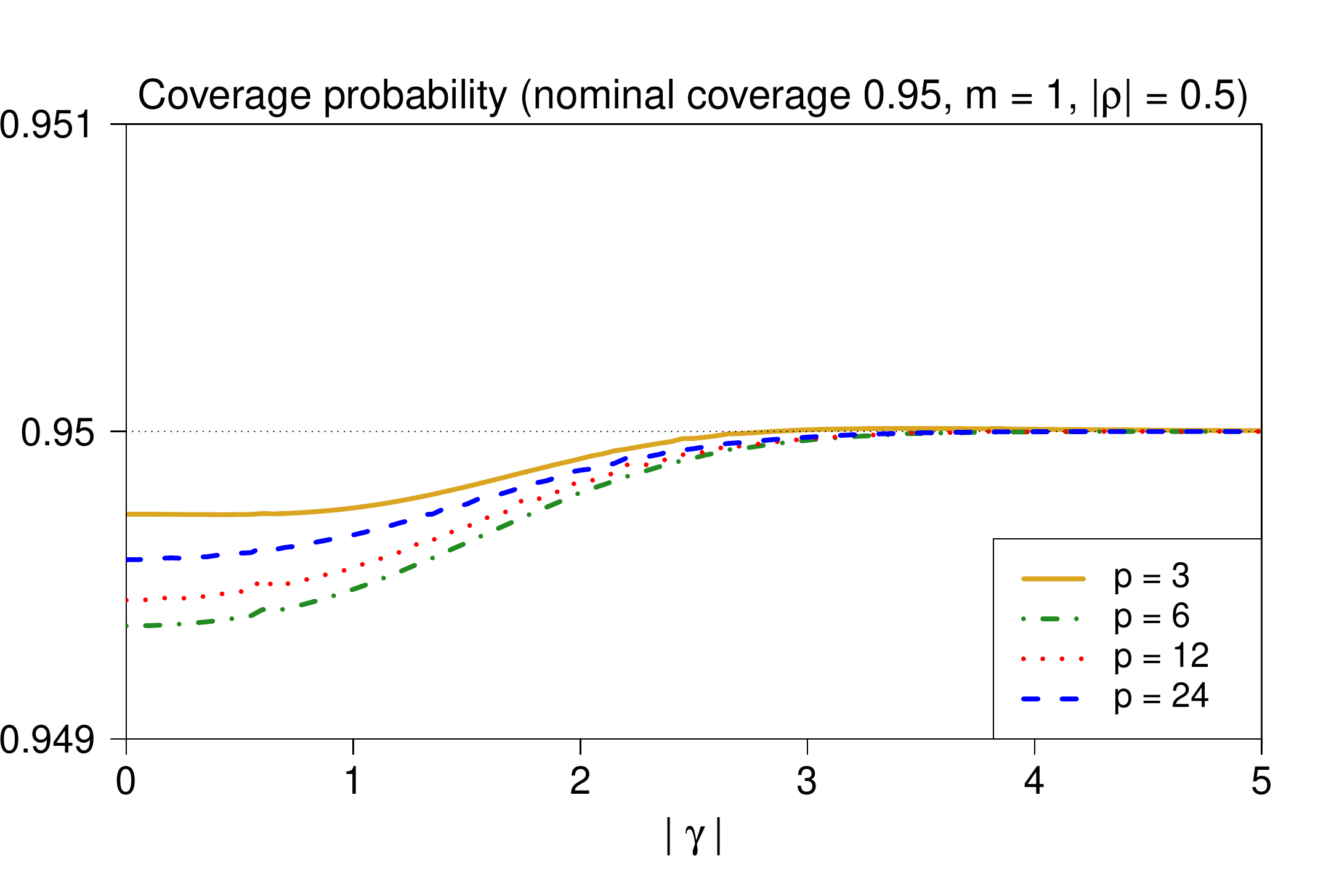}
	\includegraphics[width=0.9\textwidth]{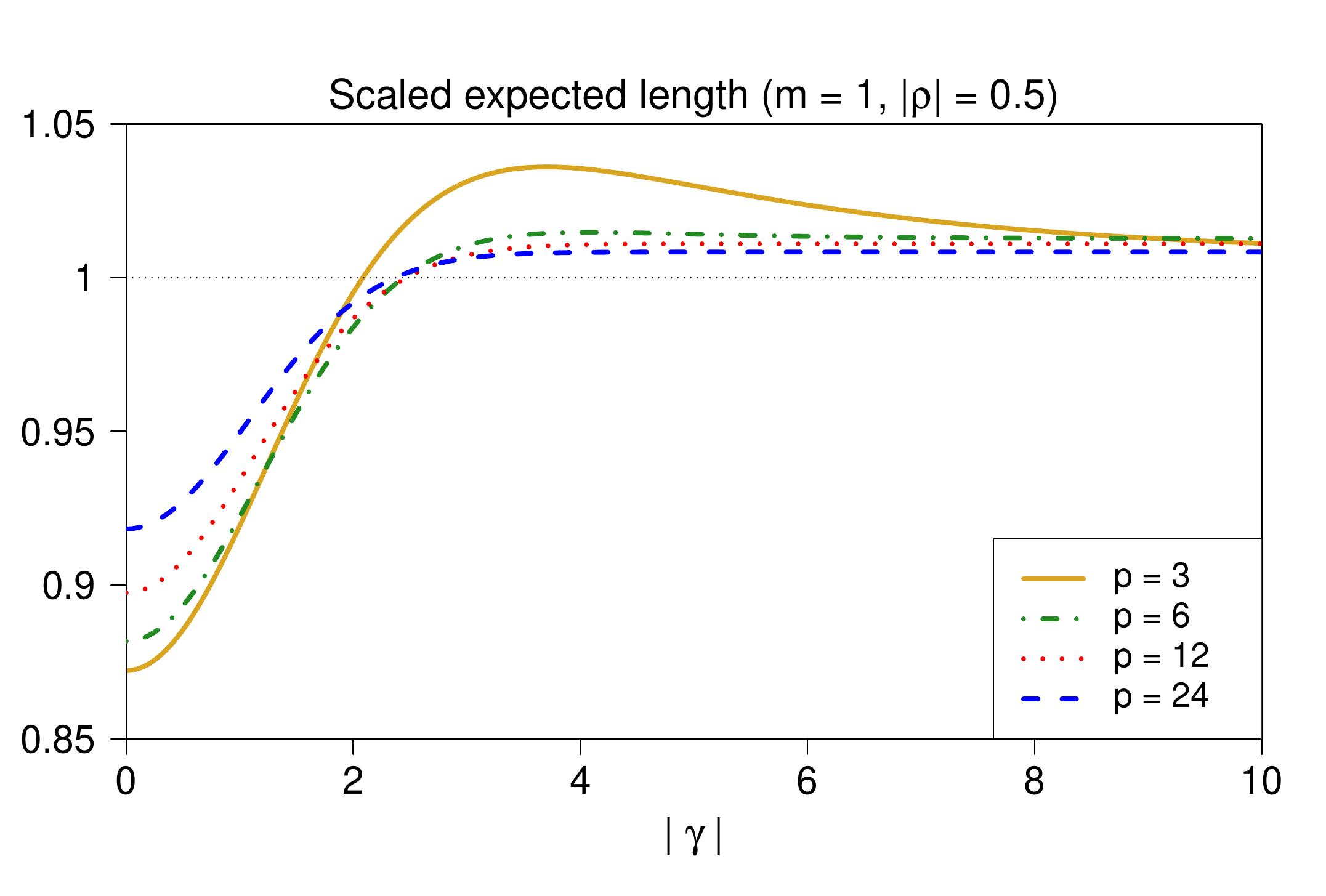}
	\caption{Coverage probability and scaled expected length of the confidence interval $J$,
		with nominal coverage 0.95, computed with AIC weights ($d=2$) for $|\rho|=0.5$ and $m=1$.}
	\label{CP_AIC_m_1_rho_2_5_p3}
\end{figure}

\begin{figure}[H]
	\centering
	\includegraphics[width=0.9\textwidth]{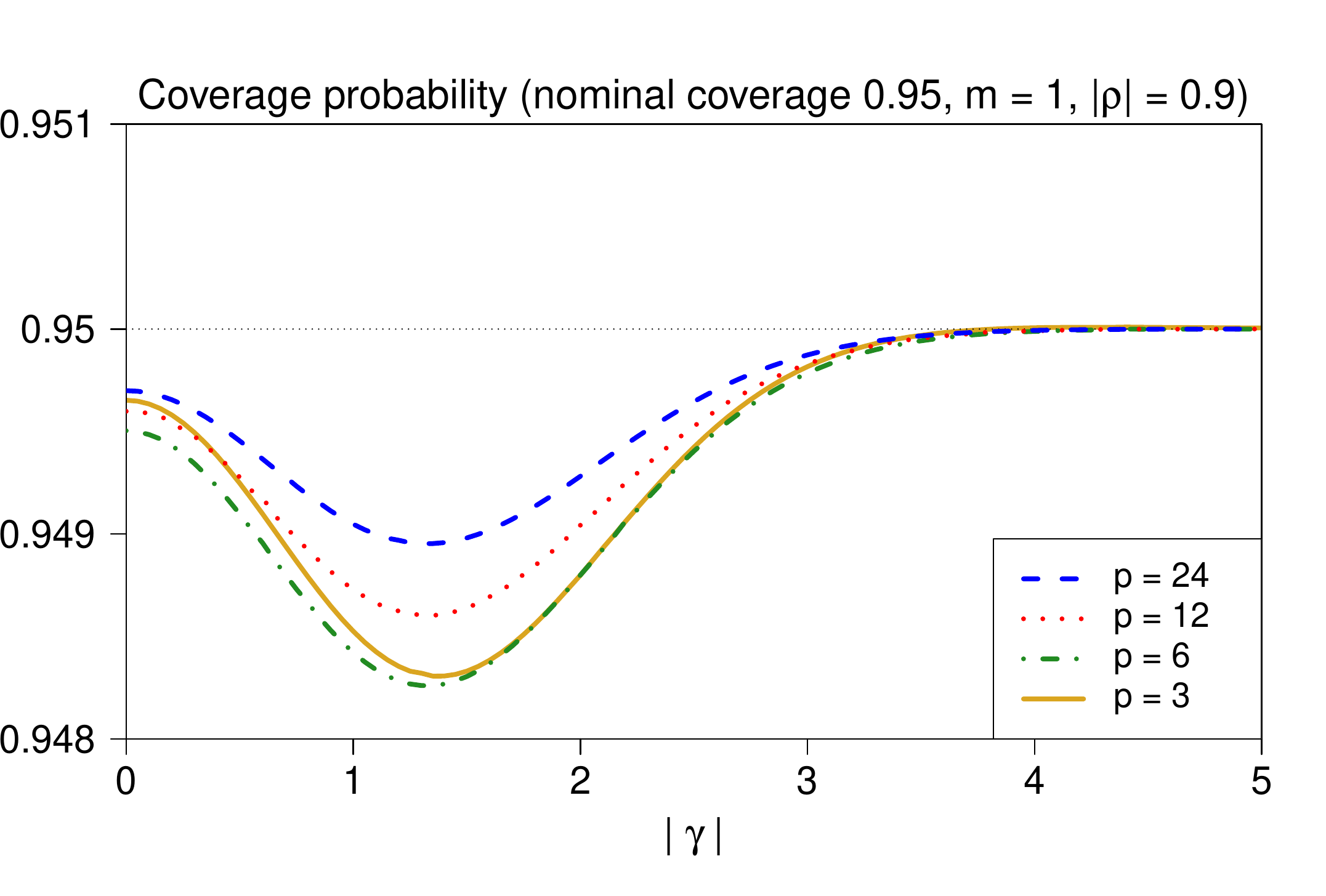}
	\includegraphics[width=0.9\textwidth]{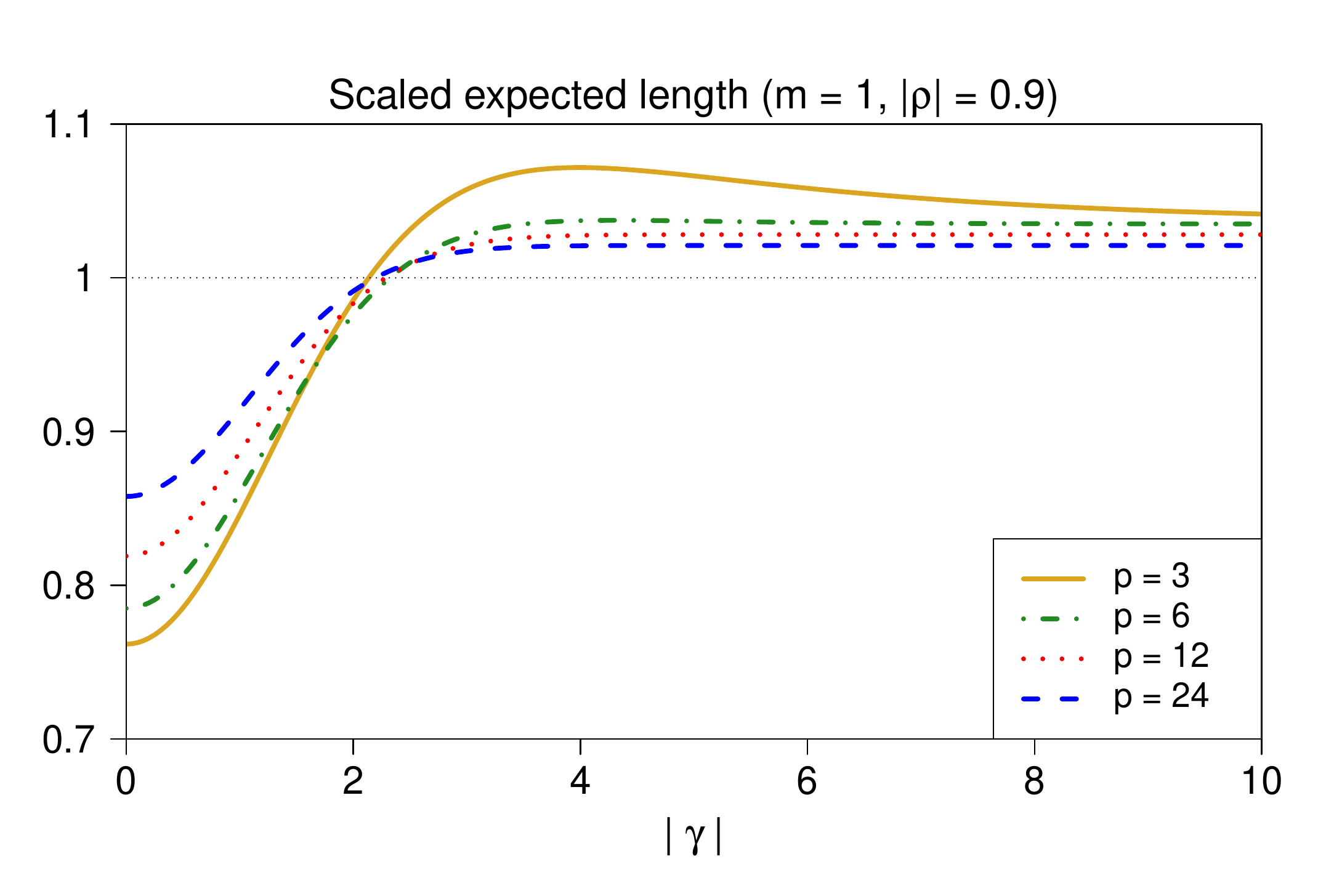}
	\caption{Coverage probability and scaled expected length of the  confidence interval $J$,
		with nominal coverage 0.95, computed with AIC weights ($d=2$) for $|\rho|=0.9$ and $m=1$.}
	\label{CP_AIC_m_1_rho_7_9_p3}
\end{figure}

\begin{figure}[H]
	\centering
	\includegraphics[width=0.9\textwidth]{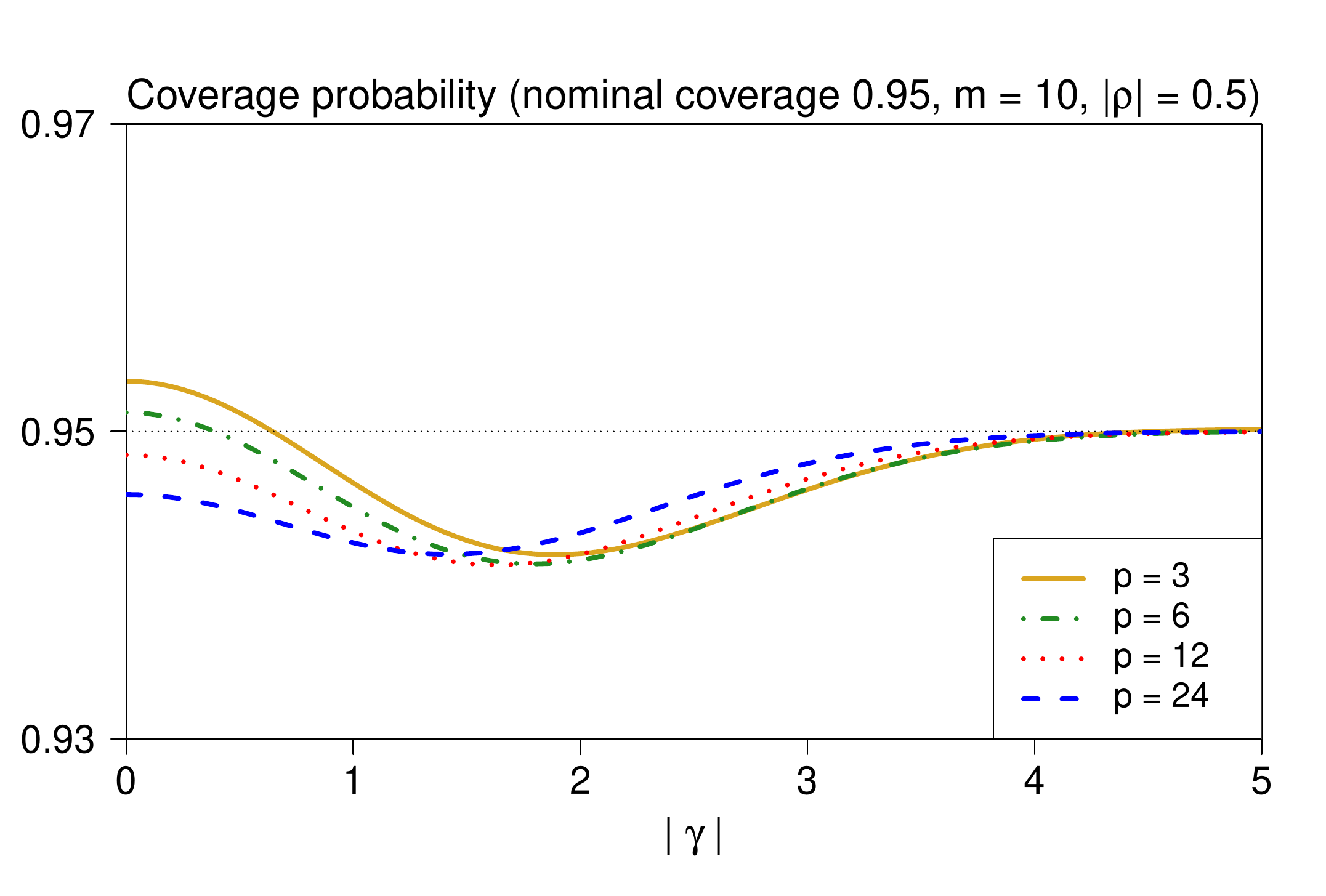}
	\includegraphics[width=0.9\textwidth]{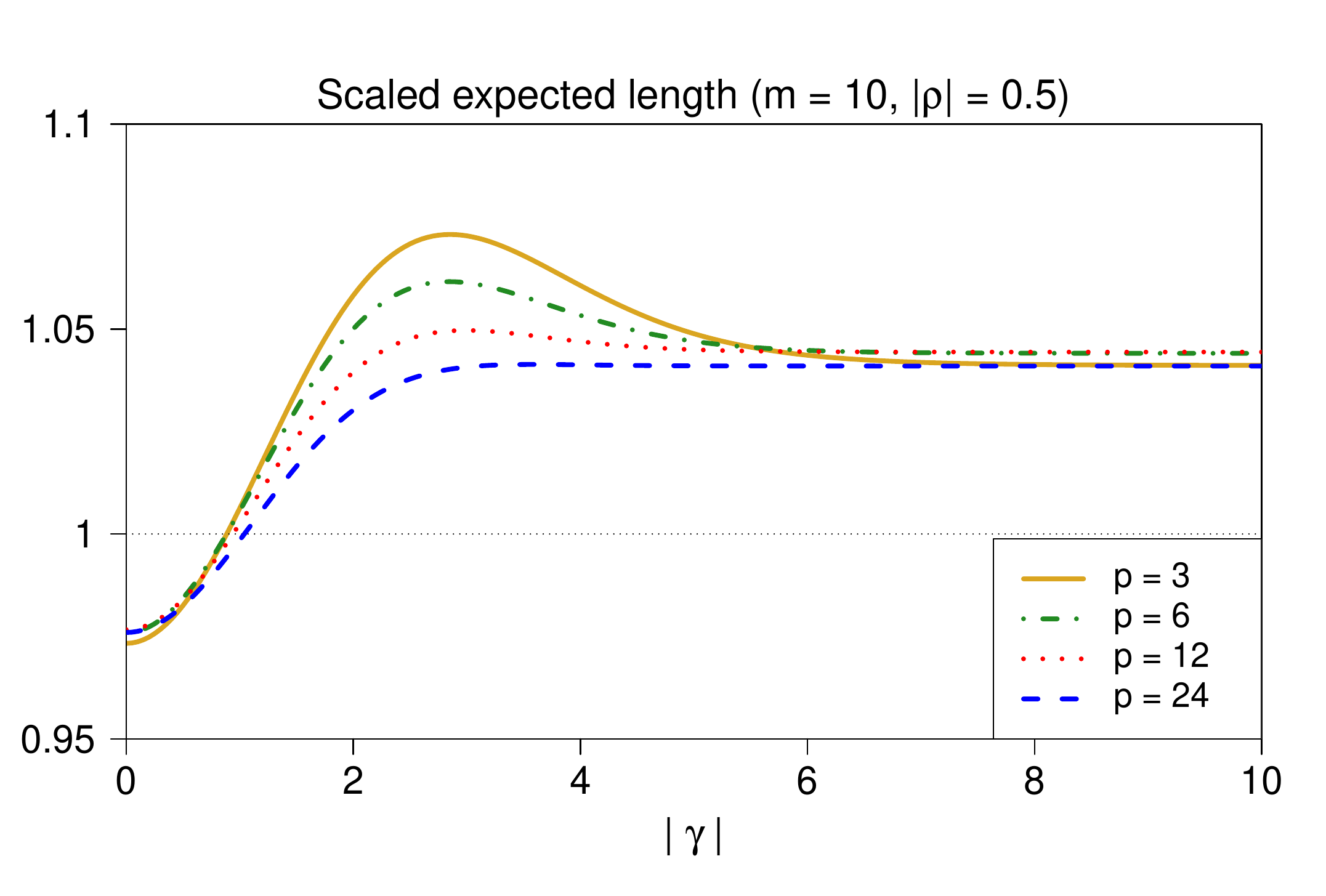}
	\caption{Coverage probability and scaled expected length of the  confidence interval $J$,
		with nominal coverage 0.95, computed with AIC weights ($d=2$) for $|\rho|=0.5$ and $m=10$.}
	\label{CP_AIC_m_10_rho_2_5_p3}
\end{figure}

\begin{figure}[H]
	\centering
	\includegraphics[width=0.9\textwidth]{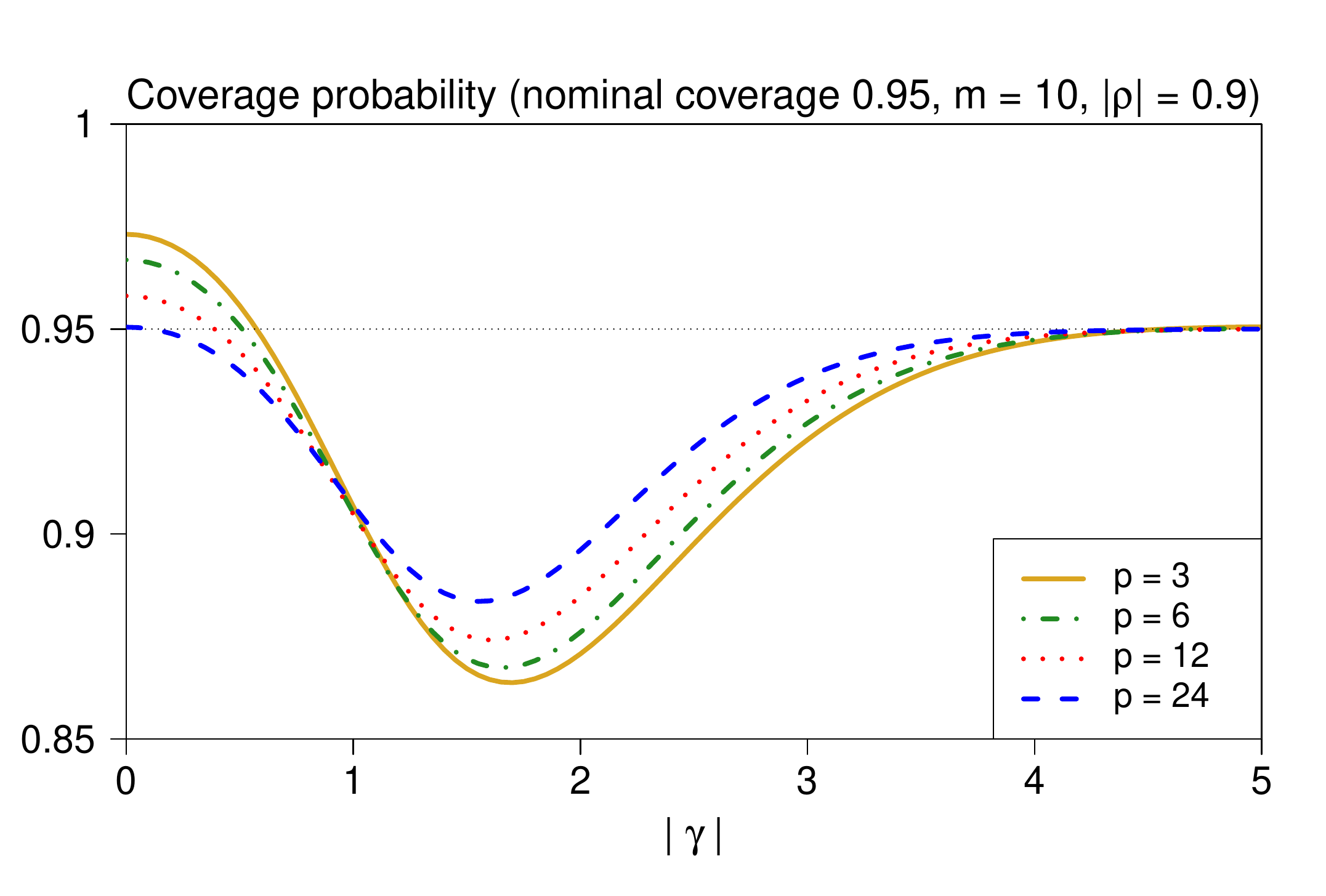}
	\includegraphics[width=0.9\textwidth]{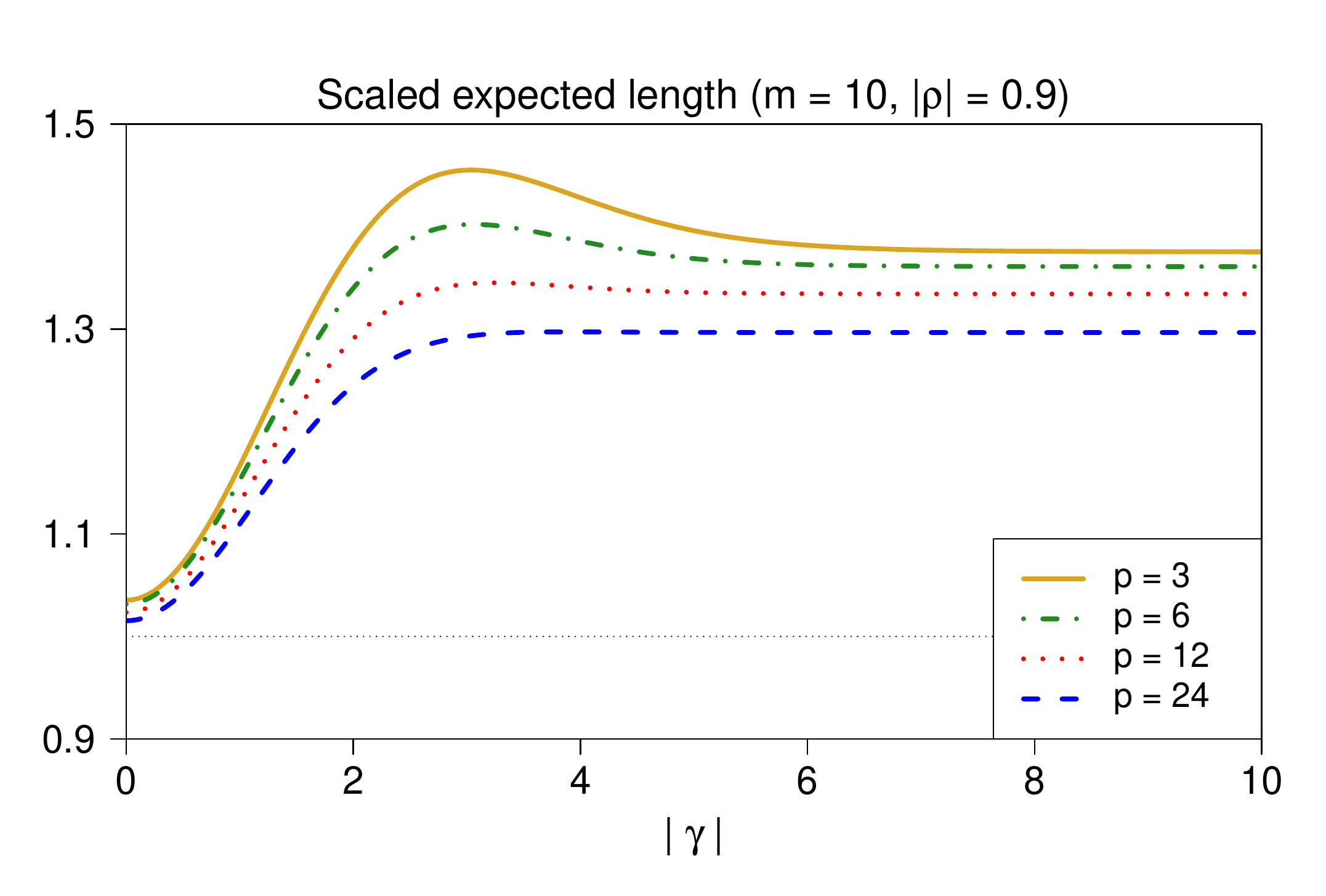}
	\caption{Coverage probability and scaled expected length of the  confidence interval $J$,
		with nominal coverage 0.95, computed with AIC weights ($d=2$) for $|\rho|=0.9$ with $m=10$.}
	\label{CP_AIC_m_10_rho_7_9_p3}
\end{figure}

Consider Figures \ref{CP_AIC_m_1_rho_2_5_p3} and \ref{CP_AIC_m_1_rho_7_9_p3} for $m=1$ and for $|\rho| = 0.5$ and $0.9$, respectively.  The minimum coverage probability of $J$ is very close to the nominal coverage $0.95$.  The scaled expected length of $J$ is substantially less than 1 when $\gamma=0$ and, although the scaled expected length of $J$ does not converge to 1 as $|\gamma| \rightarrow \infty$, the maximum value of the scaled expected length of $J$ is not too much larger than $1$.  The results for different $p \in \{ 3,6,12,24 \}$ are similar.  In these cases, when $m=1$, the model averaged confidence interval $J$ has good properties.  Figures \ref{CP_AIC_m_10_rho_2_5_p3} and \ref{CP_AIC_m_10_rho_7_9_p3} for $m=10$ and for $|\rho| = 0.5$ and $0.9$, respectively, show that the minimum coverage probability of $J$ is much lower than the nominal coverage $0.95$ and the scaled expected length of $J$ can be much larger than $1$.  The scaled expected length of $J$ has a maximum value that is an increasing function of $|\rho|$, that can be much larger than $1$ for $|\rho|$ large and $m$ not small.   That is, the performance of the confidence interval $J$ deteriorates as $m$ increases.

%Fig14b and Fig 23b

%Fig 15b and Fig 24b

%Fig 20b and Fig 29b

%Fig 21b and Fig 30b

\section{The case that $\boldsymbol{p \ge 3}$ is fixed and $\boldsymbol{m=n-p \rightarrow \infty}$} \label{largem}

Suppose that $p$ is a fixed integer, satisfying $p \ge 3$, and let $n \rightarrow \infty$, so that $m \rightarrow \infty$.  We describe the limiting behaviour of the confidence interval $J$ and the limits of the coverage probability and scaled expected length, as $m \rightarrow \infty$.   Fully rigorous proofs of the results are laborious and are not included in this paper; figures in the Supplementary Material confirm numerically that the stated limits hold.

Assume that $\bm{D} = \lim_{n \rightarrow \infty} \bX^\top\bX / n$ exists and is nonsingular. Recall that $v_\theta =  \ba^\top(\bX^\top\bX)^{-1}\ba$ and 
\begin{equation*}
\rho =  \dfrac{\ba^\top(\bX^\top\bX)^{-1}\bc}{\{\ba^\top(\bX^\top\bX)^{-1}\ba \; \bc^\top(\bX^\top\bX)^{-1}\bc\}^{1/2}}.
\end{equation*}
Although not made explicit in the notation, 
$v_{\theta}$ and $\rho$ are functions of $n$. 
Let 
\begin{equation*}
\bar{\rho} =  \dfrac{\ba^\top\bm{D}^{-1}\bc}{\{\ba^\top\bm{D}^{-1}\ba \; \bc^\top\bm{D}^{-1}\bc\}^{1/2}}.
\end{equation*}
Note that $v_{\theta} \downarrow 0$ and $\rho \rightarrow \bar{\rho}$
as $n \rightarrow \infty$. 

\subsection{Limiting behaviour of the confidence interval $\boldsymbol{J}$}

Let
\begin{equation*}
w_1^*(x) = \frac{1}{\displaystyle 1 + \exp\left( \frac{x^2 - d}{2} \right)}
\end{equation*}
and $k^*(x) = x \, w_1^*(x)$. Also let 
\begin{equation*}
r^*(x, \bar{\rho}) = w_1^*(x) \Big( 1 - \bar{\rho}^2 + \big(1 - w_1^*(x)\big)^2 \bar{\rho}^2\, x^2 \Big)^{1/2} + \big( 1 - w_1^*(x) \big) \Big( 1 + \big(w_1^*(x)\big)^2 \bar{\rho}^2 x^2 \Big)^{1/2}.
\end{equation*}
Finally, let
\begin{equation*}
J^* = \left[\widehat\theta -  \bar{\rho}\, \sigma \, v_\theta^{1/2} \, 
k^*\big(\widetilde{\gamma} \big) \,
\pm z_{1-\alpha/2} \, \sigma \, v_\theta^{1/2} \, r^* \big(\widetilde{\gamma}, \bar{\rho} \big)\right].
\end{equation*}
This interval describes the limiting behaviour of the confidence interval $J$ in
the sense described below.

For each fixed $x \in \mathbb{R}$,
\begin{align*}
w_1(x) 
= \frac{1}{\displaystyle 1 + \left( 1 + \frac{x^2/2}{m/2} \right)^{m/2} \left( 1 + \frac{x^2}{m} \right)^{p/2} \exp\big(-d/2\big) } 
\rightarrow  w_1^*(x) 
\end{align*}
 as $m \rightarrow \infty$.
%because 
%%
%\begin{equation*}
%\lim_{n \rightarrow \infty} \left( 1 + \frac{x}{n} \right)^n = \exp(x) 
%\qquad \text{and} \qquad 
%\lim_{n \rightarrow \infty} \left( 1 + \frac{x^2}{m} \right)^{p/2} = 1.
%\end{equation*}
%%
Hence, for each fixed $x \in \mathbb{R}$, 
$k(x) = x \, w_1(x) \rightarrow k^*(x)$ as $m \rightarrow \infty$.
Finally, 
for each fixed $x \in \mathbb{R}$, $r(x, \rho) \rightarrow r^*(x, \bar{\rho})$ as
$m \rightarrow \infty$.

 We compare the differences between the centres and half-widths
of $J$ and $J^*$ with $\sigma \, v_{\theta}^{1/2}$, the standard deviation of 
$\widehat{\theta}$.
Since $t_{m, 1-\alpha/2} \rightarrow z_{1 - \alpha/2}$ and 
$W \xrightarrow{p} 1$, as $m \rightarrow \infty$, the 
differences between the centers and half-widths of $J$ and $J^*$,
divided by $\sigma \, v_{\theta}^{1/2}$ converge in probability to 0 as $m \rightarrow \infty$.

\subsection{Limiting behaviour of the coverage probability}

It follows from the proof of Theorem 1 that the coverage probability 
\begin{equation*}
P\big( \theta \in J \big) = P \Big( \ell\big(\widetilde{\gamma}, W, \rho\big) \le G \le u\big(\widetilde{\gamma}, W, \rho\big) \Big),
\end{equation*}
where $G = (\widehat{\theta} - \theta)/(\sigma\, v_\theta^{1/2})$ and
\begin{align*}
\ell(\widetilde{\gamma}, W, \rho) 
&= \rho \,  W \,  k\big(\widetilde{\gamma} / W \big) - t_{m, 1-\alpha/2} \,   W \, r \big(\widetilde{\gamma} / W, \rho \big)  
\\
u(\widetilde{\gamma}, W, \rho) 
&= \rho \,  W \,  k\big(\widetilde{\gamma} / W \big) + t_{m, 1-\alpha/2} \,   W \, r \big(\widetilde{\gamma} / W, \rho \big).
\end{align*}
Let 
\begin{align*}
\ell^*(\widetilde{\gamma}, \rho) 
&= \rho \,    k^*\big(\widetilde{\gamma} \big) - z_{1-\alpha/2} \,    r^* \big(\widetilde{\gamma} , \rho \big)  
\\
u^*(\widetilde{\gamma}, \rho) 
&= \rho \,    k^*\big(\widetilde{\gamma}\big) + z_{1-\alpha/2} \,    r^* \big(\widetilde{\gamma} , \rho \big).
\end{align*}
Thus
\begin{align*}
\ell\big(\widetilde{\gamma}, W, \rho \big) - \ell^*\big( \widetilde{\gamma}, \rho \big)
\xrightarrow{p} 0 \ \text{ and } \ u\big(\widetilde{\gamma}, W, \rho \big)  - u^*\big( \widetilde{\gamma}, \rho \big) \xrightarrow{p} 0, \text{  as } m \rightarrow \infty,
\end{align*}
so that
\begin{equation*}
%\label{CP_m_infty_1}
P\big( \theta \in J \big) - P \Big( \ell^*\big( \widetilde{\gamma}, \rho \big) \le G \le u^*\big( \widetilde{\gamma}, \rho \big) \Big)  \rightarrow 0 \ \text{ as } \ m \rightarrow \infty.
\end{equation*}
Since the distribution of $G$ conditional on $\widetilde{\gamma} = h$ is $N\big( \rho(h-\gamma), 1-\rho^2 \big)$, 
\begin{align*}
&P \Big( \ell^*\big( \widetilde{\gamma}, \rho \big) \le G \le u^*\big( \widetilde{\gamma}, \rho \big) \Big) \\
&= \int_{-\infty}^{\infty} P \Big( \ell^*\big( h, \rho \big) \le G \le u^*\big( h, \rho \big) \, \Big|\,  \widetilde\gamma = h \Big) \phi(h-\gamma)\, dh\\
&= \int_{-\infty}^{\infty} P \Big( \ell^*\big( h, \rho \big) \le \widetilde{G} \le u^*\big( h, \rho \big) \Big) \phi(h-\gamma)\, dh, \quad \text{ where } \widetilde{G} \sim N\big( \rho(h-\gamma), 1-\rho^2 \big),\\
&= \int_{-\infty}^{\infty} \Psi\Big( \ell^*(h, \rho), u^*(h, \rho); \rho(h-\gamma), 1-\rho^2 \Big) \phi(h-\gamma)\, dh \\
&= \int_{-\infty}^{\infty} \Psi\Big( \ell^*(y+\gamma, \rho), u^*(y+\gamma, \rho); \rho \, y, 1-\rho^2 \Big) \phi(y)\, dy\\
&= CP^*(\gamma, \rho),
\end{align*}
say.  Consequently, 
$P\big( \theta \in J \big) \rightarrow CP^*(\gamma, \bar{\rho})$  as $m \rightarrow \infty$.

\subsection{Limiting behaviour of the scaled expected length}

We have shown that the scaled expected length of the confidence interval $J$ is 
\begin{equation}
\label{SEL_final1}
\frac{t_{m, 1-\alpha/2}}{t_{m, (1+c_{\rm min})/2}}\, \left(\frac{m}{2}\right)^{1/2} \frac{\Gamma(m/2)}{\Gamma\big((m+1)/2\big)}\, \int_{0}^{\infty} \int_{-\infty}^{\infty} w\, r\left(\frac{y+\gamma}{w}, \rho \right)\, \phi(y)\, dy \, f_W(w)\, dw.
\end{equation}
It follows from 6.1.47 of \citet[p.257]{AbramowitzStegun1964} that
\begin{equation*}
\left(\frac{m}{2}\right)^{1/2} \frac{\Gamma(m/2)}{\Gamma\big((m+1)/2\big)} \longrightarrow 1 \quad \text{ as } \quad m \rightarrow \infty.
\end{equation*}
Also, $t_{m, 1-\alpha/2} \rightarrow z_{1-\alpha/2}$ and $t_{m, (1+c_{\rm min})/2} - z_{(1 + c_{\rm min}^*)/2} \rightarrow 0$, as $m \rightarrow \infty$, where $c_{\rm min}^*$ is $CP^*(\gamma,\bar{\rho})$ minimized with respect to $\gamma \ge 0$.
In addition, it is plausible that 
\begin{equation*}
\int_{0}^{\infty} \int_{-\infty}^{\infty} w\, r\left(\frac{y+\gamma}{w}, \rho \right)\, \phi(y)\, dy \, f_W(w)\, dw - \int_{-\infty}^{\infty}  r^*\big(y+\gamma, \rho \big)\, \phi(y)\, dy \rightarrow 0, 
\end{equation*}
 as $m \rightarrow \infty$.
Therefore the difference between the scaled expected length of the confidence interval $J$ and  
\begin{equation*}
\frac{z_{1-\alpha/2}}{z_{(1+c_{\rm min}^*)/2}} \int_{-\infty}^{\infty}  r^*\big(y+\gamma, \rho \big)\, \phi(y)\, dy = SEL^*(\gamma, \rho),
\end{equation*}
say, approaches 0 as $m \rightarrow \infty$.
 Consequently, 
the scaled expected length of $J$ converges to $SEL^*(\gamma, \bar{\rho})$ as $m \rightarrow \infty$.

\subsection{Some numerical results for large $m$} 

For $\bar{\rho} = 0$, the interval $J^*$ reduces to the standard $1-\alpha$ confidence
interval for $\theta$ based on the full model, assuming that $\sigma^2$ is known.  
Consistently with this fact, for $\bar{\rho} = 0$, $CP^*(\gamma, \bar{\rho}) = 1-\alpha$ and 
$SEL^*(\gamma, \bar{\rho}) = 1$ for all $\gamma$. As $|\bar{\rho}|$ increases, $CP^*(\gamma, \bar{\rho})$
and $SEL^*(\gamma, \bar{\rho})$ increasingly differ from these values.

We first did some calculations to explore empirically the reasonableness of the limiting results stated above.  In particular, for the confidence interval $J$, with nominal coverage 0.95 and constructed using AIC weights ($d=2$), we constructed figures showing the coverage probability and the scaled expected length in the case $p=4$ and $|\rho| = 0.2, 0.5, 0.7, 0.9$ at different values of $m = 10,50,200,\infty$.  These figures (included in the Supplementary Material) show the convergence of the coverage probability and scaled expected length to the limits stated above as $m$ increases.  For all the values of $|\rho|$ we considered, the exact results for $m=200$ are very close to the limiting results,  indicating that the asymptotic results are useful at this value.   These figures also show that, other than for  small $|\rho|$, the performance of the confidence interval $J$ deteriorates in terms of both coverage probability and scaled expected length as $m \rightarrow \infty$.
 
\begin{figure}[H]
	\centering
	\includegraphics[width=0.9\textwidth]{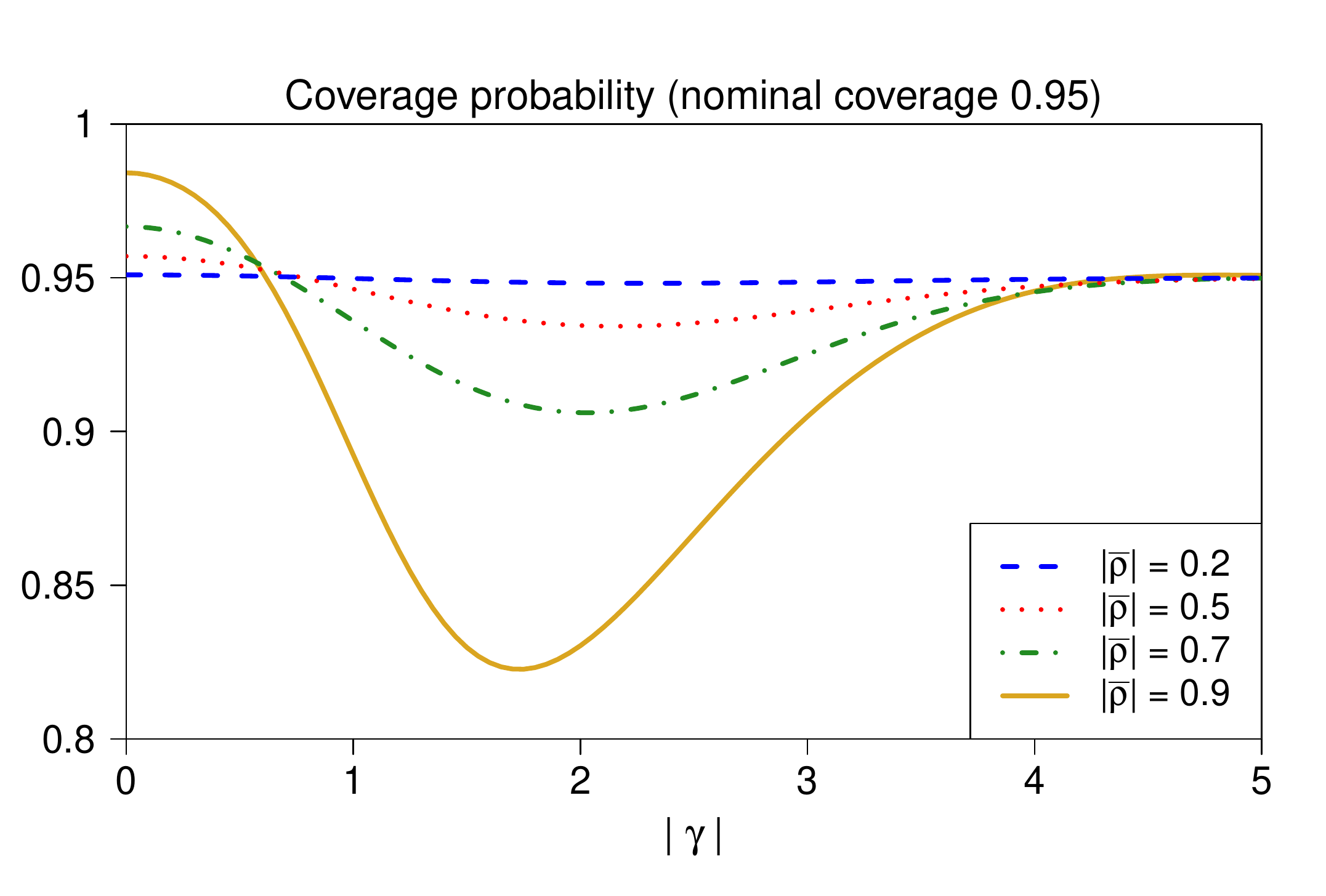}
	\includegraphics[width=0.9\textwidth]{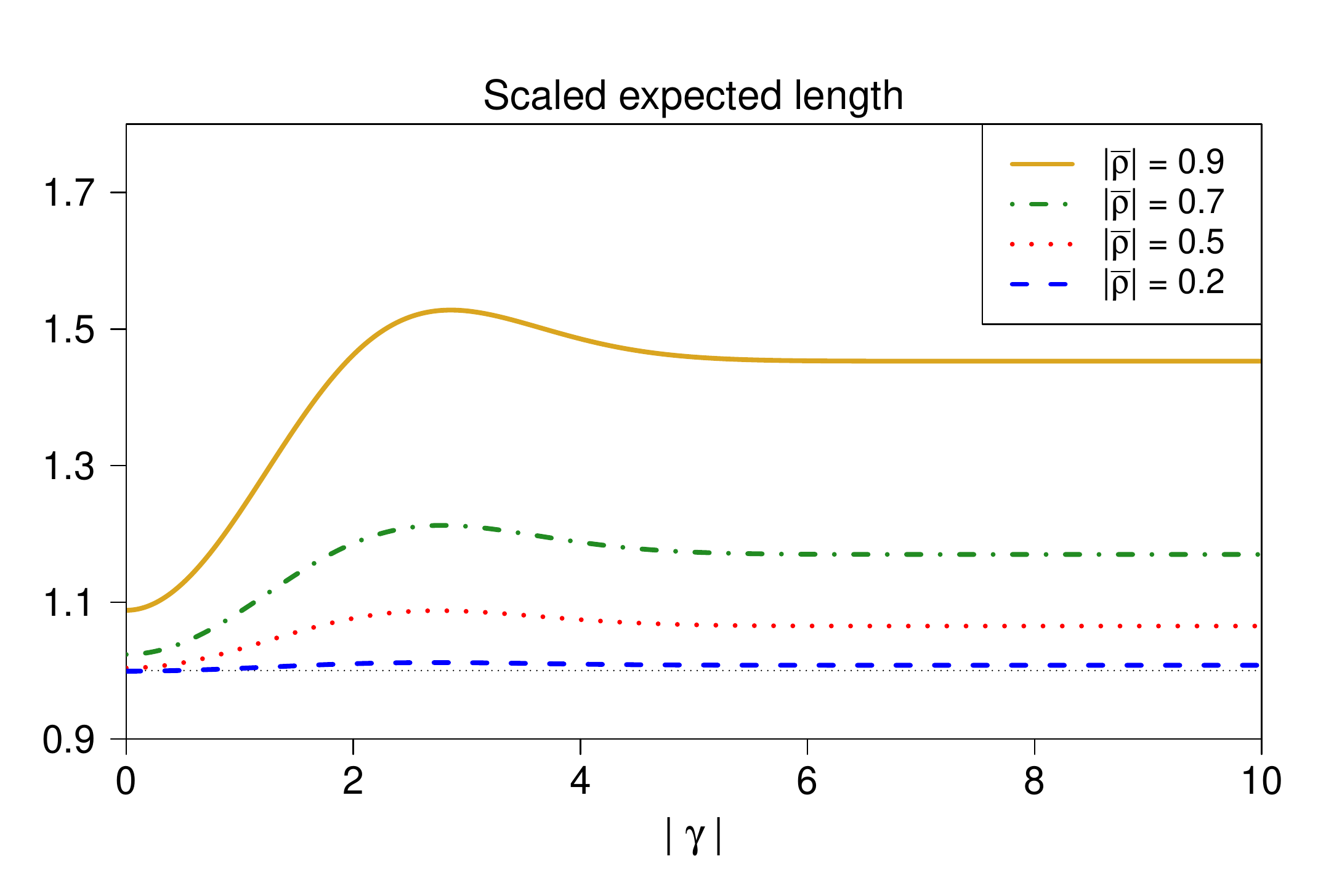}
	\caption{Coverage probability and scaled expected length of the  confidence interval $J$, with nominal coverage 0.95 and computed with AIC weights ($d=2$), when  $m \rightarrow \infty$ with $p \ge 3$ fixed.}
	\label{CP_AIC_m_inf}
\end{figure}

For the confidence interval $J$, with nominal coverage 0.95 and constructed using AIC weights ($d=2$), we present the coverage probability and the scaled expected length (Figure \ref{CP_AIC_m_inf}) in the limiting case $m \rightarrow \infty$ with $p$ fixed ($p \ge 3$) at different values of $|\bar{\rho}|=0.2, 0.5, 0.7, 0.9$.  These figures quantify how the performance of $J$ deteriorates with increasing $|\bar{\rho}|$. As expected, for small $|\bar{\rho}|$, the asymptotic coverage is the same as the nominal coverage and the scaled expected length is 1. However, with large $|\bar{\rho}|$, the minimum asymptotic coverage of the  confidence interval,
with nominal coverage 0.95, can be as low as 0.83, even though the scaled expected length is well above $1$ for all $|\gamma|$.

% m = Inf

\subsection{Comparison with asymptotic results of \cite{HjortClaeskens2003}} \label{compare}

\begin{figure}[H]
	\centering
	\includegraphics[width=0.9\textwidth]{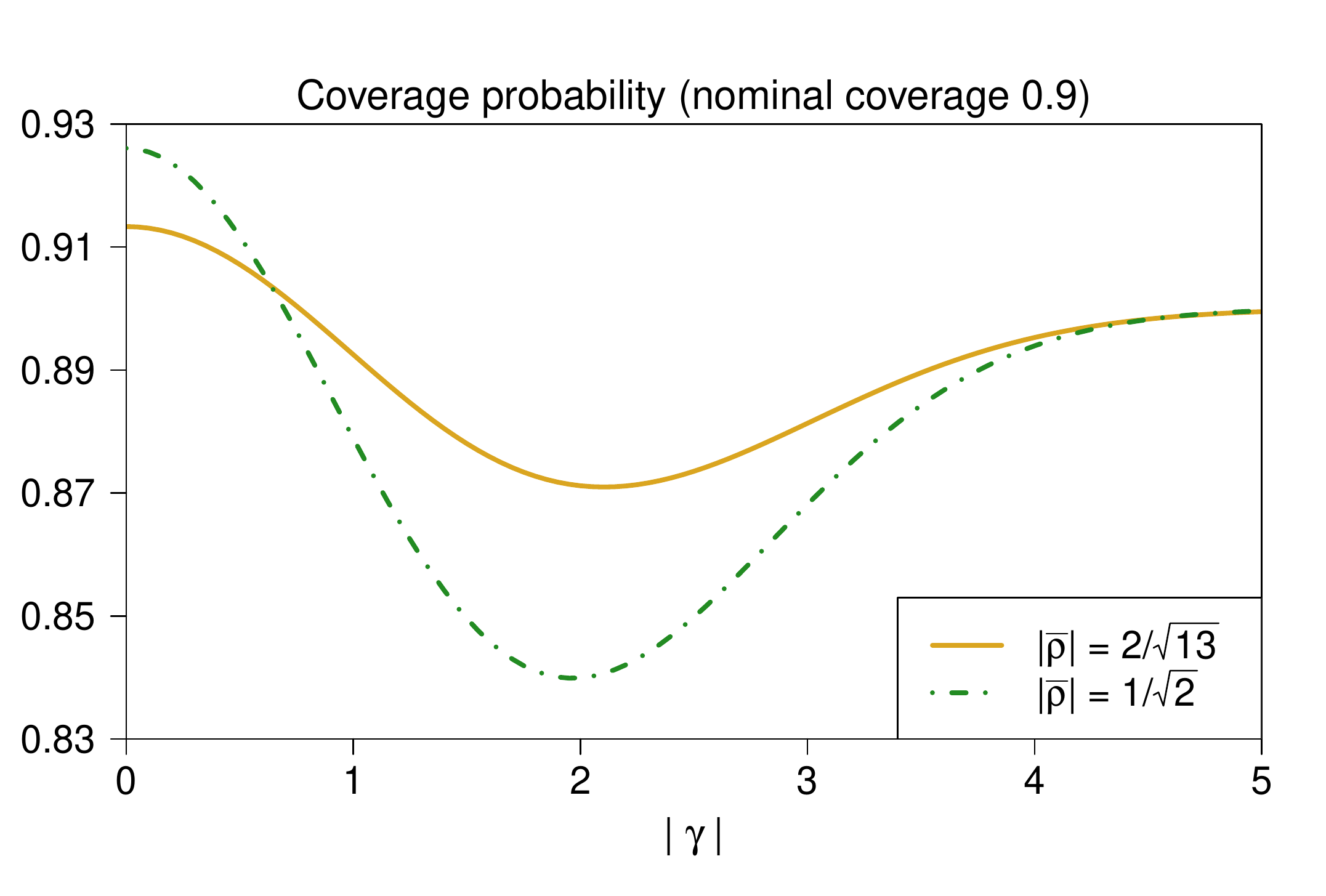}
	\caption{Coverage probability of the confidence interval $J$, with nominal coverage 0.9, computed with AIC weights ($d=2$) for $|\bar{\rho}|=2/\sqrt{13}$ (equivalent to $\rho_{HC} = 2/3$ in Figure 2 of \cite{HjortClaeskens2003}) and $|\bar{\rho}|=1/\sqrt{2}$ (equivalent to $\rho_{HC} = 1$ in Figure 2 of \cite{HjortClaeskens2003}) with $m\rightarrow \infty$ when $p \ge 3$ is fixed.}
	\label{CP_AIC_m_200_compare_Hjort}
\end{figure}

\citet[p. 886]{HjortClaeskens2003} consider two nested general regression models: the full model (which they call the extended model) and the simpler model (which they call the narrow model), where the simpler model is obtained from the full model by setting a scalar parameter to a given value.
In the solid line curves in their Figure 2, \citet[p. 886]{HjortClaeskens2003} present the limiting coverage of the \cite{BucklandEtAl1997} confidence interval using the standard error based on formula (9) of \cite{BucklandEtAl1997} in the following context.
 They consider two situations corresponding to two values of a parameter that they denote by $\rho$ and which, to avoid confusion, we will denote by $\rho_{HC}$.  In the caption of Figure 2, \cite{HjortClaeskens2003} define
\begin{equation*}
\rho_{HC} = \frac{\omega \, K^{1/2}}{\tau_0},
\end{equation*}   
with $K$ defined at the start of their Section 3.1, $\omega$ defined in their equation (3.2) and $\tau_0^2$ just below their equation (4.3).  
In the 
 Supplementary Material it is shown that $\rho_{HC}$, expressed in our notation,
is $-\bar{\rho} \big/(1 - \bar{\rho}^2)^{1/2}$.
Note that when $\rho_{HC} = 2/3$ and $\rho_{HC} = 1$, our $|\bar{\rho}|$ is equal to $2/\sqrt{13}$ and $1/\sqrt{2}$, respectively. Figure \ref{CP_AIC_m_200_compare_Hjort} shows the coverage probability of $J$ using our computations in the same situations as those considered by \cite{HjortClaeskens2003}; we observe that this  our Figure \ref{CP_AIC_m_200_compare_Hjort} is identical to 
the solid line curves of Figure 2 of \cite{HjortClaeskens2003}.

%------------------------------------------------------------------------------

\section{Discussion} \label{discuss}

In the context of a simple testbed situation involving two linear regression models, we have derived exact expressions for the coverage probability and scaled expected length of the confidence interval centered on a frequentist model averaged estimator proposed by \cite{BucklandEtAl1997}.  Using these expressions to explore the exact finite sample performance of the Buckland-Burnham-Augustin confidence interval, we showed that the confidence interval with residual degrees of freedom $m=1$ has good coverage and scaled expected properties and that these deteriorate as $m$ increases, being already quite poor for $m=10$.  We also explored the limiting asymptotic case (as $m \rightarrow \infty$) and showed that the minimum limiting coverage can be much lower than the nominal value even when the maximum scaled expected length is much larger than one, throughout the parameter space.  Differences in generality and notation mean that it is not obvious how our limiting coverage results relate to those of \cite{HjortClaeskens2003} (who did not include any results on expected length).  We were able to compare our results to those obtained for the asymptotic coverage of the confidence interval by \cite{HjortClaeskens2003} and show that they are the same.  Our results enhance the coverage result obtained by \cite{HjortClaeskens2003} by providing exact results in the more limited testbed situation for any sample size for both coverage and scaled expected length.
All the results taken together show that the Buckland-Burnham-Augustin confidence interval cannot be generally recommended.

\bigskip

\noindent \textbf{Acknowledgment}

\noindent This work was supported by an Australian Government Research Training Program Scholarship.

\baselineskip=24pt

\appendix

\section{Appendix}

\subsection{Proof of Theorem \ref{Theorem_CP}}
\label{Proof_CP}

\noindent (a)
The coverage probability of the confidence interval $J$ is
\begin{align*}
&P(\theta \in J) 
\\
&= P\left( \widetilde{\theta} - t_{m, 1-\alpha/2} \, \sigma \, W \,  v_\theta^{1/2} \, r \big(\widetilde{\gamma} / W, \rho \big) \leq  \theta \leq  \widetilde{\theta} + t_{m, 1-\alpha/2} \, \sigma \, W \,  v_\theta^{1/2} \, r \big(\widetilde{\gamma} / W, \rho \big) \right)
\\
&= P\left(  - t_{m, 1-\alpha/2} \, \sigma \, W \,  v_\theta^{1/2} \, r \big(\widetilde{\gamma} / W, \rho \big)
\leq  \widehat\theta -\theta - \rho \, v_\theta^{1/2} \, \sigma\,  W \,  k\big(\widetilde{\gamma} / W \big)  
\leq  t_{m, 1-\alpha/2} \, \sigma \, W \,  v_\theta^{1/2} \, r \big(\widetilde{\gamma} / W, \rho \big)\right)
\\
&= P\left(  - t_{m, 1-\alpha/2} \,   W \, r \big(\widetilde{\gamma} / W, \rho \big) \leq  \frac{\widehat\theta -\theta}{\sigma\, v_\theta^{1/2}} - \rho \,  W \,  k\big(\widetilde{\gamma} / W \big) 
\leq   t_{m, 1-\alpha/2} \,   W \, r \big(\widetilde{\gamma} / W, \rho \big)\right)
\\
&= P\Big( \rho \,  W \,  k\big(\widetilde{\gamma} / W \big) - t_{m, 1-\alpha/2} \,   W \, r \big(\widetilde{\gamma} / W, \rho \big)  
\leq  G
\leq \rho \,  W \,  k\big(\widetilde{\gamma} / W \big) +  t_{m, 1-\alpha/2} \,   W \, r \big(\widetilde{\gamma} / W, \rho \big)\Big), 
\end{align*}
where $G = (\widehat{\theta} - \theta)/(\sigma\, v_\theta^{1/2})$. Note that
\begin{equation*}
\left[ \begin{array}{c}
G\\
\widetilde{\gamma}
\end{array} \right] \sim N \left( 
\left[ \begin{array}{c}
0\\
\gamma 
\end{array} \right], \, \left[ \begin{array}{c c}
1 & \rho \\
\rho & 1 
\end{array}\right] \right),
\end{equation*}
so the distribution of $G$ conditional on $\widetilde{\gamma}=h$ is 
$N \big(\rho(h-\gamma), 1-\rho^2 \big)$. Recall that 
\begin{align*}
\ell(\widetilde{\gamma}, W, \rho) 
&= \rho \,  W \,  k\big(\widetilde{\gamma} / W \big) - t_{m, 1-\alpha/2} \,   W \, r \big(\widetilde{\gamma} / W, \rho \big)  
\\
u(\widetilde{\gamma}, W, \rho) 
&= \rho \,  W \,  k\big(\widetilde{\gamma} / W \big) +  t_{m, 1-\alpha/2} \,   W \, r \big(\widetilde{\gamma} / W, \rho \big).
\end{align*}
Therefore the coverage probability is
\begin{align}
\notag
CP(\gamma, \rho) &= P \Big( \ell(\widetilde{\gamma}, W, \rho) \leq G \leq u(\widetilde{\gamma}, W, \rho) \Big)
\\
\notag
&= \int_{0}^{\infty} \int_{-\infty}^{\infty} 
P \Big( \ell(\widetilde{\gamma}, W, \rho) \leq G \leq u(\widetilde{\gamma}, W, \rho) \, \Big| \, \widetilde{\gamma} = h, W=w \Big) \, \phi(h-\gamma)\, dh\, f_W(w)\, dw
\\
\notag
&= \int_{0}^{\infty} \int_{-\infty}^{\infty} 
P \Big( \ell(h, w, \rho) \leq G \leq u(h, w, \rho) \, \Big| \, \widetilde{\gamma} = h \Big) \, \phi(h-\gamma)\, dh\, f_W(w)\, dw
\\
\notag
&= \int_{0}^{\infty} \int_{-\infty}^{\infty} 
P \Big( \ell(h, w, \rho) \leq \widetilde{G} \leq u(h, w, \rho) \Big) \, \phi(h-\gamma)\, dh\, f_W(w)\, dw,
\\
\notag
& \qquad \qquad \qquad \qquad \qquad \qquad \text{where } \widetilde{G} \sim N \big(\rho(h-\gamma), 1-\rho^2 \big),
\\
\label{CP_formula1}
&= \int_{0}^{\infty} \int_{-\infty}^{\infty} \Psi \Big( \ell(h, w, \rho), u(h, w, \rho); \, \rho(h-\gamma), 1-\rho^2 \Big) \, \phi(h-\gamma)\, dh\, f_W(w)\, dw,
\end{align}
where $\Psi(a, b; \mu, v) = P(a \leq Z \leq b)$ for $Z \sim N(\mu, v)$. 
Now, by changing the variable of integration in the inner integral to $y = h-\gamma$, we obtain \eqref{CP_final}.   \hfill \qed

\noindent (b)
We use the following lemmas.

\begin{lemma}
\label{lem1}
$\Psi(a,b ; \mu, v) = \Psi(-b, -a ; -\mu, v)$, 
where $\Psi(a, b; \mu, v) = P(a \le Z \le b)$ for $Z \sim N(\mu, v)$.
\end{lemma}

\noindent Lemma \ref{lem1} is same as the Lemma 2 of \cite{KabailaWijethunga2019}. The following lemma has some similarities to Lemma 3 of \cite{KabailaWijethunga2019}.

\begin{lemma}
\label{lem2}
\begin{enumerate}[label=(\roman*)]
	\item $-u(-h, w, \rho) = \ell(h, w, \rho)$
	\item $-\ell(-h, w, \rho) = u(h, w, \rho)$
	\item $\ell(h, w, -\rho) = -u(h, w, \rho)$
	\item $u(h, w, -\rho) = -\ell(h, w, \rho)$
\end{enumerate}
\end{lemma}

\noindent {\bf Proof of Lemma \ref{lem1}:}   For  $Z \sim N(\mu, v)$,
\begin{equation*}
\Psi(a, b; \mu, v) = P(a \le Z \le b)
= P(-b \le -Z \le -a)
= \Psi(-b, -a; -\mu, v).
\end{equation*}
 \hfill \qed

\noindent {\bf Proof of Lemma \ref{lem2}:}
Remember that
\begin{align*}
\ell(h, w, \rho) &= - t_{m, 1-\alpha/2} \,   w \, r \big(h / w, \rho \big)
+ \rho \,  w \,  k\big(h / w \big) \\
u(h, w, \rho) &= t_{m, 1-\alpha/2} \,   w \, r \big(h / w, \rho \big)
+ \rho \,  w \,  k\big(h / w \big). 
\end{align*}
where
\begin{align*}
r(x, \rho) &=   w_1(x) \left( \frac{m + x^2}{m+1} (1-\rho^2) + (1-w_1(x))^2 \rho^2\, x^2 \right)^{1/2} + (1-w_1(x)) \Big( 1  + w_1^2(x) \, \rho^2\,  x^2 \Big)^{1/2},
\\
w_1(x) &= \frac{1}{1 + \left( 1 + \displaystyle 
	\frac{ x^2}{m} \right)^{n/2} \, \exp\big(-d/2 \big)}\quad \text{and} \quad 
k(x) = x \, w_1(x).
\end{align*}

\noindent Obviously, $w_1(x)$ is an even function and $k(x)$ is an odd function. Note that $r(x, \rho)$ is an even function of $x$, for given $\rho$, and an even function of $\rho$, for given $x$.

\begin{enumerate}[label=(\roman*)]
	\item Since $k$ is an odd function and $r(x, \rho)$ is an even function of $x$,
	\begin{align*}
	-u(-h, w, \rho) &=- t_{m, 1-\alpha/2} \,   w \, r \big(-h / w, \rho \big)
	- \rho \,  w \,  k\big(-h / w \big)\\
	&= - t_{m, 1-\alpha/2} \,   w \, r \big(h / w, \rho \big)
	+ \rho \,  w \,  k\big(h / w \big), \\ 
	&= \ell(h,w, \rho).
	\end{align*}
	
	\item Since $k$ is an odd function and $r(x, \rho)$ is an even function of $x$,
	\begin{align*}
	-\ell(-h, w, \rho) &= t_{m, 1-\alpha/2} \,   w \, r \big(-h / w, \rho \big)
	- \rho \,  w \,  k\big(-h / w \big)\\
	&=  t_{m, 1-\alpha/2} \,   w \, r \big(h / w, \rho \big)
	+ \rho \,  w \,  k\big(h / w \big), \\ 
	&= u(h,w, \rho).
	\end{align*}
	
	\item Since $r(x, \rho)$ is an even function of $\rho$,
	\begin{align*}
	\ell(h, w, -\rho) &= -t_{m, 1-\alpha/2} \,   w \, r \big(h / w, -\rho \big)
	- \rho \,  w \,  k\big(h / w \big)\\
	&=  - t_{m, 1-\alpha/2} \,   w \, r \big(h / w, \rho \big)
	- \rho \,  w \,  k\big(h / w \big), \\ 
	&= -u(h,w, \rho).
	\end{align*}
	
	\item Since $r(x, \rho)$ is an even function of $\rho$,
	\begin{align*}
	u(h, w,- \rho) &= t_{m, 1-\alpha/2} \,   w \, r \big(h / w, -\rho \big)
	- \rho \,  w \,  k\big(h / w \big)\\
	&=  t_{m, 1-\alpha/2} \,   w \, r \big(h / w, \rho \big)
	- \rho \,  w \,  k\big(h / w \big), \\ 
	&= -\ell(h,w, \rho).
	\end{align*}
	\hfill \qed
\end{enumerate}

\noindent From \eqref{CP_formula1} 
\begin{align*}
CP(\gamma, \rho) = \int_{0}^{\infty} \int_{-\infty}^{\infty} \Psi \Big( \ell(h, w, \rho), u(h, w, \rho); \, \rho(h-\gamma), 1-\rho^2 \Big) \, \phi(h-\gamma)\, dh\, f_W(w)\, dw.
\end{align*}
Consider
\begin{align*}
CP(-\gamma, \rho) &= \int_{0}^{\infty} \int_{-\infty}^{\infty} \Psi \Big( \ell(h, w, \rho), u(h, w, \rho); \, \rho(h+\gamma), 1-\rho^2 \Big) \, \phi(h+\gamma)\, dh\, f_W(w)\, dw\\
&= \int_{0}^{\infty} \int_{-\infty}^{\infty} \Psi \Big( -u(h, w, \rho), -\ell(h, w, \rho); \, -\rho(h+\gamma), 1-\rho^2 \Big) \, \phi(-h-\gamma)\, dh\, f_W(w)\, dw,
\end{align*}
from Lemma \ref{lem1} and since $\phi$ is an even function. Now by changing the variable of integration in the inner integral to $y = -h$, 
\begin{align*}
CP(-\gamma, \rho) &= \int_{0}^{\infty} \int_{-\infty}^{\infty} \Psi \Big( -u(-y, w, \rho), -\ell(-y, w, \rho); \, \rho(y-\gamma), 1-\rho^2 \Big) \, \phi(y-\gamma)\, dy\, f_W(w)\, dw \\
&= \int_{0}^{\infty} \int_{-\infty}^{\infty} \Psi \Big( \ell(y, w, \rho), u(y, w, \rho); \, \rho(y-\gamma), 1-\rho^2 \Big) \, \phi(y-\gamma)\, dy\, f_W(w)\, dw, \\
&\qquad \qquad \qquad \qquad \qquad \qquad \text{from Lemma \ref{lem2}(i) and Lemma \ref{lem2}(ii),}\\
&= CP(\gamma, \rho).
\end{align*}
Therefore, $CP(\gamma, \rho)$ is an even function of $\gamma$, for given $\rho$.

Now consider
\begin{align*}
CP(\gamma, -\rho) &= \int_{0}^{\infty} \int_{-\infty}^{\infty} \Psi \Big( \ell(h, w, -\rho), u(h, w, -\rho); \, -\rho(h-\gamma), 1-\rho^2 \Big) \, \phi(h-\gamma)\, dh\, f_W(w)\, dw\\
&= \int_{0}^{\infty} \int_{-\infty}^{\infty} \Psi \Big( -u(h, w, \rho), -\ell(h, w, \rho); \, -\rho(h-\gamma), 1-\rho^2 \Big) \, \phi(h-\gamma)\, dh\, f_W(w)\, dw,\\
&\qquad \qquad \qquad \qquad \text{from Lemma \ref{lem2}(iii) and Lemma \ref{lem2}(iv)},\\
&= \int_{0}^{\infty} \int_{-\infty}^{\infty} \Psi \Big( \ell(h, w, \rho), u(h, w, \rho); \, \rho(h-\gamma), 1-\rho^2 \Big) \, \phi(h-\gamma)\, dh\, f_W(w)\, dw,\\
&\qquad \qquad \qquad \qquad \text{from Lemma \ref{lem1}},\\
&= CP(\gamma, \rho).
\end{align*}
Therefore, $CP(\gamma, \rho)$ is an even function of $\rho$, for given $\gamma$.  \hfill \qed

\subsection{Proof of Theorem \ref{Theorem_SEL}}
\label{Proof_SEL}

\noindent (a)
The scaled expected length is
\begin{align*}
\frac{E \Big( \text{length of the confidence interval $J$} \Big)}{E \Big( \text{length of the standard CI with the same coverage as the minimum coverage of $J$} \Big)}.
\end{align*}
Observe that
\begin{align*}
E \Big( \text{length of the confidence interval $J$} \Big) &= E \Big(2\, t_{m, 1-\alpha/2} \, \sigma \, W \,  v_\theta^{1/2} \, r \big(\widetilde{\gamma} / W, \rho \big) \Big)\\
&= 2\, t_{m, 1-\alpha/2} \, \sigma \, v_\theta^{1/2}\, E \Big( W\, r\big(\widetilde{\gamma}/ W, \rho \big) \Big).
\end{align*}
Let $c_{\rm min}$ be the minimum coverage probability of the confidence interval $J$. Then the standard confidence interval with coverage $c_{\rm min}$ is 
$\big[\widehat{\theta} \pm t_{m, (1+c_{\rm min})/2} \, \widehat\sigma \, v_\theta^{1/2}\big]$. Thus
\begin{align*}
&E \Big( \text{length of the standard CI with the same coverage as the minimum coverage of $J$} \Big)\\
&\qquad \qquad \qquad = E \Big( 2\, t_{m, (1+c_{\rm min})/2} \, \widehat\sigma \, v_\theta^{1/2}  \Big)\\
&\qquad \qquad \qquad = 2\, t_{m, (1+c_{\rm min})/2} \, \sigma \, v_\theta^{1/2} E(W).
\end{align*}
Therefore the scaled expected length
\begin{equation*}
SEL(\gamma) = \frac{t_{m, 1-\alpha/2}}{t_{m, (1+c_{\rm min})/2}}\, \frac{E \Big( W\, r\big(\widetilde{\gamma}/ W, \rho \big) \Big)}{E (  W )}.
\end{equation*}

\medskip

\noindent Note that $W \sim \big( Q/m \big)^{1/2}$ where $Q \sim \chi^2_m$. Therefore
\begin{align*}
E(W) = E \left( \frac{Q^{1/2}}{m^{1/2}} \right)
= \frac{1}{m^{1/2}} \, 2^{1/2}\, \frac{\Gamma\big((m+1)/2\big)}{\Gamma(m/2)}
= \left(\frac{m}{2}\right)^{-1/2} \, \frac{\Gamma\big((m+1)/2\big)}{\Gamma(m/2)}.
\end{align*}
Thus
\begin{align}
\notag
SEL(\gamma, \rho) &= \frac{t_{m, 1-\alpha/2}}{t_{m, (1+c_{\rm min})/2}}\, \left(\frac{m}{2}\right)^{1/2} \frac{\Gamma(m/2)}{\Gamma\big((m+1)/2\big)}\,E \Big( W\, r\big(\widetilde{\gamma}/ W, \rho \big) \Big)\\
\label{SEL_formula1}
&= \frac{t_{m, 1-\alpha/2}}{t_{m, (1+c_{\rm min})/2}}\, \left(\frac{m}{2}\right)^{1/2} \frac{\Gamma(m/2)}{\Gamma\big((m+1)/2\big)}\, \int_{0}^{\infty} \int_{-\infty}^{\infty} w\, r(h/w, \rho)\, \phi(h-\gamma)\, dh \, f_W(w)\, dw.
\end{align}
Now, by changing the variable of integration in the inner integral to $y = h-\gamma$, we obtain \eqref{SEL_final}.  

\hfill \qed  

\noindent (b)
From \eqref{SEL_formula1}
\begin{align*}
SEL(\gamma, \rho) = \frac{t_{m, 1-\alpha/2}}{t_{m, (1+c_{\rm min})/2}}\, \left(\frac{m}{2}\right)^{1/2} \frac{\Gamma(m/2)}{\Gamma\big((m+1)/2\big)}\, \int_{0}^{\infty} \int_{-\infty}^{\infty} w\, r(h/w, \rho)\, \phi(h-\gamma)\, dh \, f_W(w)\, dw.
\end{align*}
We known that $r(x, \rho)$ is an even function of $x$, for given $\rho$, and an even function of $\rho$, for given $x$. Consider the inner integral 
\begin{align*}
SEL_1(\gamma, \rho) &= \int_{-\infty}^{\infty}  r(h/w, \rho)\, \phi(h-\gamma)\, dh,
\end{align*}
which depends on $\gamma$ and $\rho$.
If we can show that $SEL_1(\gamma, \rho)$ is an even function of $\gamma$, for given $\rho$, and an even function of $\rho$, for given $\gamma$, then we can say that $SEL(\gamma, \rho)$ is also an even function of $\gamma$, for given $\rho$, and an even function of $\rho$, for given $\gamma$. Now consider
\begin{align*}
SEL_1(-\gamma, \rho) &= \int_{-\infty}^{\infty}  r(h/w, \rho)\, \phi(h+\gamma)\, dh\\
&= \int_{-\infty}^{\infty}  r(h/w, \rho)\, \phi(-h-\gamma)\, dh, \quad \text{since $\phi$ is an even function,}\\
&= \int_{-\infty}^{\infty}  r(-y/w, \rho)\, \phi(y-\gamma)\, dy, \\
&\qquad \qquad \qquad \qquad \text{by changing the variable of integration to $y=-h$,}\\
&= \int_{-\infty}^{\infty}  r(y/w, \rho)\, \phi(y-\gamma)\, dy, \quad \text{since } r(x, \rho) \text{ is an even function of } x,\\
&= SEL_1(\gamma, \rho).
\end{align*}
Therefore $SEL_1(\gamma, \rho)$ is an even function of $\gamma$, for given $\rho$. Thus $SEL(\gamma, \rho)$ is an even function of $\gamma$, for given $\rho$.
Consider
\begin{align*}
SEL_1(\gamma, -\rho) &= \int_{-\infty}^{\infty}  r(h/w, -\rho)\, \phi(h-\gamma)\, dh\\
&= \int_{-\infty}^{\infty}  r(h/w, \rho)\, \phi(h-\gamma)\, dh, \quad \text{since } r(x, \rho) \text{ is an even function of } \rho,\\
&= SEL_1(\gamma, \rho).
\end{align*}
Therefore $SEL_1(\gamma, \rho)$ is an even function of $\rho$, for given $\gamma$. Thus $SEL(\gamma, \rho)$ is an even function of $\rho$, for given $\gamma$.  \hfill \qed

\end{document}